\documentclass[ALICE,manyauthors]{cernphprep}
\usepackage[comma,square,numbers,sort&compress]{natbib}
\usepackage{hyperref}
\usepackage{lineno}
\usepackage{xspace}
\usepackage{xcolor}
\usepackage{mathtools}
\usepackage{multirow}
\usepackage{amsmath}
\usepackage{graphicx}
\usepackage{csquotes}
\usepackage[T1]{fontenc}
\usepackage{orcidlink}

\begin{document}
%

\newcommand{\pp}           {pp\xspace}
\newcommand{\ppbar}        {\mbox{$\mathrm {p\overline{p}}$}\xspace}
\newcommand{\XeXe}         {\mbox{Xe--Xe}\xspace}
\newcommand{\PbPb}         {\mbox{Pb--Pb}\xspace}
\newcommand{\pA}           {\mbox{pA}\xspace}
\newcommand{\pPb}          {\mbox{p--Pb}\xspace}
\newcommand{\AuAu}         {\mbox{Au--Au}\xspace}
\newcommand{\dAu}          {\mbox{d--Au}\xspace}

\newcommand{\s}            {\ensuremath{\sqrt{s}}\xspace}
\newcommand{\snn}          {\ensuremath{\sqrt{s_{\mathrm{NN}}}}\xspace}
\newcommand{\pt}           {\ensuremath{p_{\rm T}}\xspace}
\newcommand{\meanpt}       {$\langle p_{\mathrm{T}}\rangle$\xspace}
\newcommand{\ycms}         {\ensuremath{y_{\rm CMS}}\xspace}
\newcommand{\ylab}         {\ensuremath{y_{\rm lab}}\xspace}
\newcommand{\etarange}[1]  {\mbox{$\left | \eta \right |~<~#1$}}
\newcommand{\yrange}[1]    {\mbox{$\left | y \right |~<~#1$}}
\newcommand{\dndy}         {\ensuremath{\mathrm{d}N_\mathrm{ch}/\mathrm{d}y}\xspace}
\newcommand{\dndeta}       {\ensuremath{\mathrm{d}N_\mathrm{ch}/\mathrm{d}\eta}\xspace}
\newcommand{\avdndeta}     {\ensuremath{\langle\dndeta\rangle}\xspace}
\newcommand{\dNdy}         {\ensuremath{\mathrm{d}N_\mathrm{ch}/\mathrm{d}y}\xspace}
\newcommand{\Npart}        {\ensuremath{N_\mathrm{part}}\xspace}
\newcommand{\Ncoll}        {\ensuremath{N_\mathrm{coll}}\xspace}
\newcommand{\dEdx}         {\ensuremath{\textrm{d}E/\textrm{d}x}\xspace}
\newcommand{\RpPb}         {\ensuremath{R_{\rm pPb}}\xspace}

\newcommand{\nineH}        {$\sqrt{s}~=~0.9$~Te\kern-.1emV\xspace}
\newcommand{\seven}        {$\sqrt{s}~=~7$~Te\kern-.1emV\xspace}
\newcommand{\twoH}         {$\sqrt{s}~=~0.2$~Te\kern-.1emV\xspace}
\newcommand{\twosevensix}  {$\sqrt{s}~=~2.76$~Te\kern-.1emV\xspace}
\newcommand{\five}         {$\sqrt{s}~=~5.02$~Te\kern-.1emV\xspace}
\newcommand{\twosevensixnn}{$\sqrt{s_{\mathrm{NN}}}~=~2.76$~Te\kern-.1emV\xspace}
\newcommand{\fivenn}       {$\sqrt{s_{\mathrm{NN}}}~=~5.02$~Te\kern-.1emV\xspace}
\newcommand{\LT}           {L{\'e}vy-Tsallis\xspace}
\newcommand{\GeVc}         {Ge\kern-.1emV/$c$\xspace}
\newcommand{\MeVc}         {Me\kern-.1emV/$c$\xspace}
\newcommand{\TeV}          {Te\kern-.1emV\xspace}
\newcommand{\GeV}          {Ge\kern-.1emV\xspace}
\newcommand{\MeV}          {Me\kern-.1emV\xspace}
\newcommand{\GeVmass}      {Ge\kern-.2emV/$c^2$\xspace}
\newcommand{\MeVmass}      {Me\kern-.2emV/$c^2$\xspace}
\newcommand{\lumi}         {\ensuremath{\mathcal{L}}\xspace}

\newcommand{\ITS}          {\rm{ITS}\xspace}
\newcommand{\TOF}          {\rm{TOF}\xspace}
\newcommand{\ZDC}          {\rm{ZDC}\xspace}
\newcommand{\ZDCs}         {\rm{ZDCs}\xspace}
\newcommand{\ZNA}          {\rm{ZNA}\xspace}
\newcommand{\ZNC}          {\rm{ZNC}\xspace}
\newcommand{\SPD}          {\rm{SPD}\xspace}
\newcommand{\SDD}          {\rm{SDD}\xspace}
\newcommand{\SSD}          {\rm{SSD}\xspace}
\newcommand{\TPC}          {\rm{TPC}\xspace}
\newcommand{\TRD}          {\rm{TRD}\xspace}
\newcommand{\VZERO}        {\rm{V0}\xspace}
\newcommand{\VZEROA}       {\rm{V0A}\xspace}
\newcommand{\VZEROC}       {\rm{V0C}\xspace}
\newcommand{\Vdecay} 	   {\ensuremath{V^{0}}\xspace}

\newcommand{\ee}           {\ensuremath{e^{+}e^{-}}} 
\newcommand{\pip}          {\ensuremath{\pi^{+}}}
\newcommand{\pim}          {\ensuremath{\pi^{-}}}
\newcommand{\kap}          {\ensuremath{\rm{K}^{+}}}
\newcommand{\kam}          {\ensuremath{\rm{K}^{-}}}
\newcommand{\pbar}         {\ensuremath{\rm\overline{p}}}
\newcommand{\kzero}        {\ensuremath{{\rm K}^{0}_{\rm{S}}}}
\newcommand{\lmb}          {\ensuremath{\Lambda}\xspace}
\newcommand{\almb}         {\ensuremath{\overline{\Lambda}}\xspace}
\newcommand{\Om}           {\ensuremath{\Omega^-}\xspace}
\newcommand{\Mo}           {\ensuremath{\overline{\Omega}^+}\xspace}
\newcommand{\X}            {\ensuremath{\Xi^-}\xspace}
\newcommand{\Ix}           {\ensuremath{\overline{\Xi}^+}\xspace}
\newcommand{\Xis}          {\ensuremath{\Xi^{\pm}}\xspace}
\newcommand{\Oms}          {\ensuremath{\Omega^{\pm}}\xspace}
\newcommand{\degree}       {\ensuremath{^{\rm o}}\xspace}
\newcommand{\nsigTPC}      {\ensuremath{$|n\sigma|_{\mathrm{TPC}}$}}
\newcommand{\nsigTOF}      {\ensuremath{$|n\sigma|_{\mathrm{TOF}}$}}
\newcommand{\ENfive}            {\ensuremath{\sqrt{s_{\rm NN}} =~5.02}}
\newcommand{\ENtwosevensix}        {\ensuremath{\sqrt{s_{\rm NN}} =~2.76}}
\newcommand{\hpm}          {\ensuremath{\mathrm{ h}^{\pm}}}
\newcommand{\pipm}          {\ensuremath{\mathrm{ \pi}^{\pm}}}
\newcommand{\Kpm}          {\ensuremath{\mathrm{ K}^{\pm}}}
\newcommand{\vtwo}          {\ensuremath{ v_{2}}}
\newcommand{\vthree}          {\ensuremath{ v_{3}}}
\newcommand{\vn}          {\ensuremath{ v_{n}}}
\newcommand{\Ach}          {\ensuremath{ A_{\rm ch}}}
\newcommand{\nsigmaTPC}          {\ensuremath{ |\rm n\sigma|_{\mathrm{TPC}}}}
\newcommand{\nsigmaTOF}          {\ensuremath{ |\rm n\sigma|_{\mathrm{TOF}}}}
\newcommand{\DelIntCov}          {\ensuremath{ \Delta~\rm{Int.~Cov.}}}
\newcommand{\rnormvtwo}          {\ensuremath{ r_{2}^{\rm Norm}}}
\newcommand{\rnormvthree}          {\ensuremath{ r_{3}^{\rm Norm}}}

\begin{titlepage}
\PHyear{2023}       
\PHnumber{191}      
\PHdate{29 August}  

\title{Probing the Chiral Magnetic Wave with charge-dependent flow measurements in Pb--Pb collisions at the LHC}
\ShortTitle{Probing the CMW in Pb--Pb collisions at the LHC}   

\Collaboration{ALICE Collaboration\thanks{See Appendix~\ref{app:collab} for the list of collaboration members}}
\ShortAuthor{ALICE Collaboration} 

\begin{abstract}

The Chiral Magnetic Wave (CMW) phenomenon is essential to provide insights into the strong interaction in QCD, the properties of the quark--gluon plasma, and the topological characteristics of the early universe, offering a deeper understanding of fundamental physics in high-energy collisions. Measurements of the charge-dependent anisotropic flow coefficients are studied in Pb--Pb collisions at center-of-mass energy per nucleon--nucleon collision \snn $=$ 5.02 TeV to probe the CMW. In particular, the slope of the normalized difference in elliptic (\vtwo) and triangular (\vthree) flow coefficients of positively and negatively charged particles as a function of their event-wise normalized number difference, is reported for inclusive and identified particles. The slope \rnormvthree\xspace is found to be larger than zero and to have a magnitude similar to \rnormvtwo\xspace, thus pointing to a large background contribution for these measurements. Furthermore, \rnormvtwo\xspace can be described by a blast wave model calculation that incorporates local charge conservation. In addition, using the event shape engineering technique yields a fraction of CMW ($f_{\rm CMW}$) contribution to this measurement which is compatible with zero. This measurement provides the very first upper limit for $f_{\rm CMW}$, and in the 10--60\% centrality interval it is found to be 26\% (38\%) at 95\% (99.7\%) confidence level.

\end{abstract}
\end{titlepage}

\setcounter{page}{2} 


\section{Introduction} 
\label{Section:Introduction}

The primary goal of relativistic heavy-ion collisions at the Large Hadron Collider (LHC) is to study the properties of the emerging strongly interacting medium called the quark--gluon plasma (QGP)~\cite{Busza:2018rrf, Schukraft:1993fcq, Heinz:2008tv, Shuryak:1980tp, Collins:1974ky, ALICE:2022wpn}. The transition from normal hadronic matter to the QGP is predicted by quantum chromodynamics (QCD) calculations on the lattice~\cite{Aoki:2006we, Brown:1990ev}. Heavy-ion collisions are also characterized by extremely strong short-lived electromagnetic fields ({B $\sim10^{18}$~Gauss}), primarily induced by protons from the incoming nuclei that do not undergo any inelastic collision and are referred to as spectators~\cite{Bzdak:2011yy}. The direction of {\bf B} is perpendicular to the reaction plane, the plane spanned by the impact parameter of the colliding nuclei and the beam direction. The presence of this intense magnetic field allows for the possibility to study novel QCD phenomena, such as parity violation in strong interactions~\cite{Lee:1973iz,Lee:1974ma, Kharzeev:1998kz}.

The potential to observe parity violation in strong interactions using ultrarelativistic heavy-ion collisions was first discussed in Refs.~\cite{Morley:1983wr,Kharzeev:1998kz,Kharzeev:1999cz} and further reviewed in Refs.~\cite{Kharzeev:2013ffa,Kharzeev:2015kna,kharzeev_chiral_2021,Voloshin:2004vk}. Theoretically, the interactions of quarks with gluonic fields describing transitions between topologically different QCD vacuum states change the quark chirality, leading to a local chiral imbalance. The strong magnetic field leads to a charge separation (electric current) relative to the reaction plane, which is known as the Chiral Magnetic Effect (CME)~\cite{Kharzeev:2007tn,Fukushima:2008xe,Kharzeev:2007jp,Kharzeev:1998kz,Li:2014bha,Liu:2020ymh,Gao:2020vbh}. The experimental search for the CME using heavy-ion collisions has intensified over the past decade. Though early measurements pointed to some similarities between the results and the theoretical predictions~\cite{STAR:2009wot, STAR:2009tro, ALICE:2012nhw}, there is substantial evidence that background sources, i.e., collective phenomena and local charge conservation (LCC), play a significant role in the experimental measurements~\cite{Schlichting:2010qia, Pratt:2010zn}. The LCC here refers to the principle that within a local region of a physical system, the balance or conservation of quantum numbers for eg., electric charge is upheld. Experimental results indicate that the upper limit of the CME signal contribution ranges from 7\% to 20\% at 95\% confidence level in semicentral heavy-ion collisions~\cite{ALICE:2012nhw, STAR:2009wot, STAR:2009tro, Adamczyk:2013kcb, Adamczyk:2013hsi, Adamczyk:2014mzf, Acharya:2017fau, CMS:2016wfo, STAR:2019xzd, Sirunyan:2017quh, Schlichting:2010qia, Pratt:2010zn, li_chiral_2020, STAR:2019bjg, abdallah_search_2022}.

A dual phenomenon to the CME is the Chiral Separation Effect (CSE)~\cite{Son:2004tq, Metlitski:2005pr}, which is theorized to induce a chirality current along \textbf{B}  in the presence of a finite electric chemical potential ($\mu_{\rm e}$). The CME and the CSE interact with one another forming a long wavelength collective excitation, called the Chiral Magnetic Wave (CMW)~\cite{Burnier:2011bf,Burnier:2012ae,Kharzeev:2010gd,Yee:2013cya,Taghavi:2013ena}. Similar to the CME-induced electric dipole moment, the CMW would manifest itself in a finite electric quadrupole moment in the final state~\cite{Burnier:2011bf}. This effect, if present, can be measured by charge-dependent anisotropic flow measurements~\cite{Burnier:2011bf}. The anisotropic flow is quantified in terms of the Fourier coefficients $v_{\rm n}$ of the azimuthal distribution of the produced particles with respect to the $n^{\mathrm{th}}$-order event plane angle $\Psi_n$~\cite{Voloshin:1994mz, Poskanzer:1998yz, wang_number--constituent-quark_2022}
\begin{equation}
	\frac{dN}{d \varphi} \propto
	1 + \sum_{n = 1} ^\infty 2 v_{n} \cos \left[n(\varphi-\Psi_{n})\right],
	\label{eqn:flow}
\end{equation}
where $\varphi$ is the azimuthal angle of a particle. The first three coefficients $v_1$, $v_2$, and $v_3$ are known as the directed, elliptic, and triangular flow, respectively. The CMW-induced electric quadrupole moment evolves with the medium expansion, leading to an increase (decrease) of \vtwo\xspace for negatively (positively) charged hadrons~\cite{Burnier:2011bf}. The difference between negatively and positively charged hadron \vtwo\xspace($\Delta$\vtwo) is expected to be proportional to the event-by-event charge asymmetry ($A_{\rm ch}$)~\cite{Burnier:2011bf, Burnier:2012ae},
\begin{equation} 
	\Delta v_{2} = v_{2}^{-} - v_{2}^{+} \propto r_{2}A_{\rm ch}.
	\label{eq:v2ach}
\end{equation}
In the above equation, $r_{2}$ denotes the slope parameter between $\Delta v_{2}$ and event-by-event charge asymmetry, and \Ach\xspace is defined as  
\begin{equation} \label{eq:Ach}
	A_{\rm ch} = \frac{(N^{+} -N^{-})}{(N^{+} +N^{-})},
\end{equation}
where $N^{+}$ ($N^{-}$) are positively (negatively) charged hadrons measured in a given event. 

The first experimental search for the CMW was performed by the STAR Collaboration at the Relativistic Heavy Ion Collider (RHIC) with charged pions in Au--Au collisions at center-of-mass energy per nucleon--nucleon collision \snn = 200 GeV~\cite{STAR:2015wza}, in which a positive linear dependence on \Ach\xspace was observed for the \vtwo\xspace difference between \pim\xspace and \pip. The extracted positive slopes as well as their centrality dependence agree well with theoretical calculations~\cite{Burnier:2011bf,Burnier:2012ae,Kharzeev:2010gd}. 
A similar positive correlation was measured by the ALICE Collaboration at the Large Hadron Collider (LHC) with charged hadrons in semicentral Pb--Pb collisions at \snn = 2.76 TeV~\cite{ALICE:2015cjr}. Comparable slopes to those from Au--Au collisions in semicentral collisions were reported. However, the lifetime of the magnetic field in vacuum is expected to drop much faster at LHC energies compared to that at RHIC energies~\cite{Deng:2012pc}. Thus, it is highly unlikely that an identical slope value would be observed by different experiments with orders of magnitude difference in collision energies. Furthermore, a similar linear dependence was observed by the precision measurements of the CMS collaboration in p–Pb and Pb–Pb collisions at \snn = 5.02 TeV~\cite{CMS:2017pah}. This similarity questioned the CMW interpretation since the CMW signal is not expected to be present in p–Pb collisions due to the decoupling of the magnetic field from the reaction plane in such collisions~\cite{Belmont:2016oqp, CMS:2016wfo}. In addition, both STAR and CMS collaborations have observed a linear dependence between \Ach\xspace and $\Delta$\vthree\xspace, i.e. the difference between the \vthree\xspace coefficients of negatively and positively charged hadrons~\cite{STAR:2022hfy, CMS:2017pah}
\begin{equation} 
	\Delta v_{3} = v_{3}^{-} - v_{3}^{+} \propto r_{3}A_{\rm ch}.
	\label{eq:v3ach}
\end{equation}
However, this should not originate from the CMW-induced electric quadrupole configuration in the medium as the CMW is mainly driven by the magnetic field which is uncorrelated with the third order event plane. As a result of these observations, it appears likely that the slope observed in $\Delta v_{2}$ as a function of \Ach\xspace is not due to the CMW processes only. 
To ease the comparison between measurements performed by different experiments, one can define a normalized slope parameter as, 
\begin{equation} \label{eq:normDeltaVn}
	\Delta v_{n}^{\rm Norm} = \frac{v_{n}^{-} - v_{n}^{+}}{(v_{n}^{-} + v_{n}^{+})/2} \propto r_{n}^{\mathrm{Norm}} A_{\rm ch} ,
\end{equation}
where $n$ $=$ 2 or 3.
Recently, it was proposed in Ref.~\cite{Wang:2021nvh}, that one can also utilize the event shape engineering (ESE) technique~\cite{Schukraft:2012ah} to estimate the CMW signal. This selection methodology was already employed to constrain the CME~\cite{Acharya:2017fau, Sirunyan:2017quh}. The ESE approach utilizes the fluctuations in the shape of the initial state of the system and allows one to select events with the same centrality but different initial geometry, thus varying the background contributions. Instead of the \Ach--\vtwo\xspace slope, an alternative observable, the integral covariance~\cite{ALICE:2015cjr} can be used. It is defined as
\begin{equation} \label{eq:intcov}
	\Delta\rm{IC} =
	\bigg(\langle \textit{v}_{2}^{-} \textit{A}_{\rm ch}\rangle-\langle \textit{A}_{\rm ch}\rangle\langle \textit{v}_{2}^{-}\rangle\bigg) -
	\bigg(\langle \textit{v}_{2}^{+} \textit{A}_{\rm ch}\rangle-\langle \textit{A}_{\rm ch}\rangle\langle \textit{v}_{2}^{+}\rangle\bigg),
\end{equation}
where the angular bracket denotes the average over the events. This observable, by definition, calculates the covariance between \Ach\xspace and \vtwo\xspace and is equivalent to the slope parameter. The main advantage of such a covariance is the removal of the dependence on the detector acceptance and on the reconstruction efficiency of charged hadrons when expressed differentially~\cite{ALICE:2015cjr}. In addition, one no longer needs to divide each sub-sample of \vtwo\xspace into several \Ach\xspace intervals allowing for a reduction of the statistical fluctuations.

Understanding the background components and how they contribute to the experimental measurements is crucial to isolate the CMW signal. Among several background sources~\cite{Bzdak:2013yla, Stephanov:2013tga, Campbell:2013ika, Voloshin:2014gja, Hatta:2015hca, Hongo:2013cqa, Zhao:2019ybo, Xu:2019pgj}, the most dominant one is expected to be the LCC, convoluted with the collective motion of the QGP medium. The LCC mechanism depicts a scenario where pairs of particles with opposite charges are usually generated from resonance decays. Such particle production mechanism is studied with balance function measurements in heavy-ion collisions~\cite{Bass:2000az, ALICE:2013vrb}. In the CMW measurement, when one of the particles from the charge-conserving pair escapes from the limited detector acceptance, a non-zero \Ach\xspace is consequently generated~\cite{Bzdak:2013yla}. It is demonstrated in Ref.~\cite{Wu:2020wem} that the selection of specific \Ach\xspace values automatically biases the $\eta$--\pt phase space. This can trivially give rise to a \Ach--$\Delta$\vtwo\xspace correlation because of the \vtwo\xspace dependence on $\eta$ and \pt leading to non-zero slopes, even in absence of CMW phenomena. Theoretical studies on \Ach--$\Delta v_2$ correlations, without the CMW process, have been extensively investigated in Refs.~\cite{Ma:2014iva, Zhou:2018rkh, Han:2019fce, Shen:2019puh, Magdy:2020xqs}. 
The consensus is that the LCC interpretation can effectively explain both the observed \Ach--$\Delta v_3$ and \Ach--$\Delta v_2$ relations. 
A pure LCC mechanism is expected to lead to an identical~\cite{Bzdak:2013yla, CMS:2017pah} positive linear correlation between \Ach--$\Delta v_2^{\rm Norm}$ and \Ach--$\Delta v_3^{\rm Norm}$. Consequently, this implies that any difference between the normalized slopes \rnormvtwo\xspace and \rnormvthree\xspace may indicate the existence of the CMW signal.

Although there are several measurements of CMW at LHC energies, there is lack of measurements with identified hadrons. Given that the predominant background influence on CMW arises from the interplay of LCC and elliptic flow ($v_{2}$), it would be useful to measure the CMW for identified particles, as it would provide us a better handle to  control the background related to $v_{2}$~\cite{Burnier:2011bf}. 
The first theoretical study~\cite{Burnier:2011bf} predicted that only light quarks, i.e., u and d, are influenced by the chiral anomaly. However, recent theoretical calculations~\cite{Shi:2017cpu} suggest that the mass difference between the strange quark (s) and the u, d quarks can be neglected, indicating the possibility of exploring CMW effects with charged kaons. 
Nevertheless, the significant differences in the absorption cross section for (anti-)protons and kaons in the hadronic matter might obscure the signal. 
Additionally, a hydrodynamic study~\cite{Hatta:2015hca} suggests that the isospin chemical potential ($\mu_{\rm I}$) and the strangeness chemical potential ($\mu_{\rm S}$) can play essential roles. This study predicts a negative slope for kaons at RHIC energies~\cite{Hatta:2015hca}. 
Therefore, it is difficult to disentangle the CMW signal and various background contributions, if the measurements are performed only with inclusive charged hadrons.  

This paper presents the first measurement of normalized slopes \rnormvtwo\xspace and \rnormvthree\xspace for charged hadrons and identified \pipm, \Kpm, and p$+\overline{\rm p}$ in Pb--Pb collisions at \snn $=$ 5.02 TeV. These measurements will provide experimental input to the ongoing theoretical developments for the flavor dependence of the chiral anomalies. Measurements from data are further compared with a recently developed blast wave model calculation, incorporating the LCC background (BW+LCC)~\cite{wu_global_2023}. The measurement of integral covariance is also utilized to estimate an upper limit on the CMW contribution, for the first time, in Pb--Pb collisions at \snn $=$ 5.02 TeV.

This article is organized as follows: Section~\ref{Section:ExperimentalSetup} briefly describes the experimental setup, while Section~\ref{Section:Analysis} discusses the data sample, the selection criteria, and the analysis details. Section~\ref{Section:Systematics} describes the evaluation of the 
systematic uncertainties. The results are discussed and
compared with model calculations in Section~\ref{Section:Results}. A summary is outlined in Section~\ref{Section:Summary}.

\section{Experimental apparatus and data sample} 
\label{Section:ExperimentalSetup}

The ALICE detector and its performance are described in
detail in Refs.~\cite{Aamodt:2008zz, Abelev:2014ffa}. The apparatus consists of a central barrel at midrapidity ($|\eta|<0.9$), embedded in a cylindrical solenoid which provides a magnetic field of 0.5 T parallel to the beam direction, and a set of forward detectors.

Charged particles produced in the collisions at midrapidity are tracked by the Inner Tracking System (ITS)~\cite{Aamodt:2008zz} and the Time Projection Chamber (TPC)~\cite{Alme:2010ke}. The ITS, composed of the Silicon Pixel Detector (SPD), Silicon Drift Detector (SDD), and Silicon Strip Detector (SSD), consists of six cylindrical silicon layers surrounding the beam vacuum pipe. The TPC, surrounding the ITS, provides up to 159 points for track reconstruction along with specific energy loss (d$E$/d$x$) measurements, which are utilized for charged-particle identification (PID). The PID is complemented by a Time-Of-Flight (TOF) detector~\cite{ALICE:2000xcm}, which measures the flight time of charged particles. The TOF detector provides pion--kaon separation at 3$\sigma$ level up to \pt $\simeq$ 2.5 GeV/$c$ and pion--proton separation up to \pt$\simeq$ 4 GeV/$c$.

On either sides of the interaction point, the \VZERO~scintillator
arrays~\cite{Abbas:2013taa}, are used for triggering and event classification. The V0 detector consists of two arrays of 32 scintillator tiles covering the pseudorapidity ranges $2.8<\eta<5.1$ (V0A) and $-3.7<\eta<-1.7$ (V0C). Both V0 detectors are segmented in four rings in the radial direction with each ring divided into eight sectors in the azimuthal direction. The \VZEROC is also used for ESE and event selection. Two tungsten-quartz neutron Zero Degree Calorimeters (ZDCs)~\cite{Aamodt:2008zz}, positioned 112.5 meters from the interaction point on each side, are used to reduce the contamination from beam-induced background. Using the time information from V0 and ZDC, offline event selection is performed to reject background from beam--gas collisions, from parasitic beam--beam interactions, and pileup events~\cite{ALICE:2020siw, Abelev:2014ffa}. 

The analysis is performed using the data sample collected with the ALICE apparatus in the 2018 LHC Pb--Pb run at \snn = 5.02 TeV. The centrality intervals were defined in terms of percentiles of the hadronic Pb--Pb cross section, determined from selections on the sum of the V0 signal amplitudes~\cite{ALICE:2013hur}. Central and semicentral Pb--Pb collisions were selected online by applying thresholds on the V0 signal amplitudes resulting in two separate trigger classes (central and semicentral triggers).

Only events with a reconstructed primary vertex located between $\pm$ 10 cm with respect to the nominal interaction point along the beam direction ($z$ axis of the ALICE reference frame) are considered. The analysis is performed in different centrality intervals spanning from 0--5\% which corresponds to the most central collisions to 50--60\% corresponding to peripheral collisions. The total number of analyzed events after the event selection is approximately 240 million.

\section{Analysis procedure} 
\label{Section:Analysis}

Charged particles reconstructed using the \TPC~and the \ITS~information within $|\eta| <$ 0.8 and $0.2 < \pt < 10$~GeV/$c$ are considered to determine \Ach. For the measurement of flow coefficients, charged particles are restricted to $0.2 < \pt < 2.0$~GeV/$c$.
Tracks are selected requiring $|\eta|~<~\rm 0.8$, at least 70 (out of a maximum of 159) TPC space points, and $\chi^{2}$ per TPC cluster $<$ 2.5 for the momentum fit in the TPC. In order to reduce the contamination from secondary particles (i.e., particles originating from weak decays, conversions, and secondary hadronic interactions in the detector material) only tracks with a maximum distance of closest approach (DCA) to the reconstructed primary vertex in the transverse ($|\rm{DCA}_{xy}|~<$ 2.4 cm) and the longitudinal directions ($|\rm{DCA}_{z}|~<$ 3.2 cm) are accepted.
Furthermore, tracks are required to have at least one hit in the two \SPD~layers. 
Charged pions, kaons, and (anti)protons are identified from the difference between the measured and expected values of d$E$/d$x$ in TPC and time of flight to the TOF detector, expressed in units of resolution(\nsigmaTPC, \nsigmaTOF), and applying a selection on the number of accepted $\rm n\sigma$ (see Table~\ref{Tab:SystematicSources}). Additionally, tracks without TOF information with \pt larger than 0.5 GeV/$c$ for pions, 0.45 GeV/$c$ for kaons, and 0.6 GeV/$c$ for protons are rejected.
All particle species are required to lie within the rapidity range $|y| <$ 0.5. By applying these selection criteria, the efficiency of identifying charged hadrons is approximately 70\% around \pt $=$ 0.5 GeV/$c$ and increases to about 80\% for \pt values above 1 GeV/$c$. Moreover, these selection criteria guarantee a purity exceeding 90\% for all particle species across the entire range of \pt values considered in the analysis.

\begin{figure}[!hbt]
	\centering
	\includegraphics[width=0.99\textwidth]{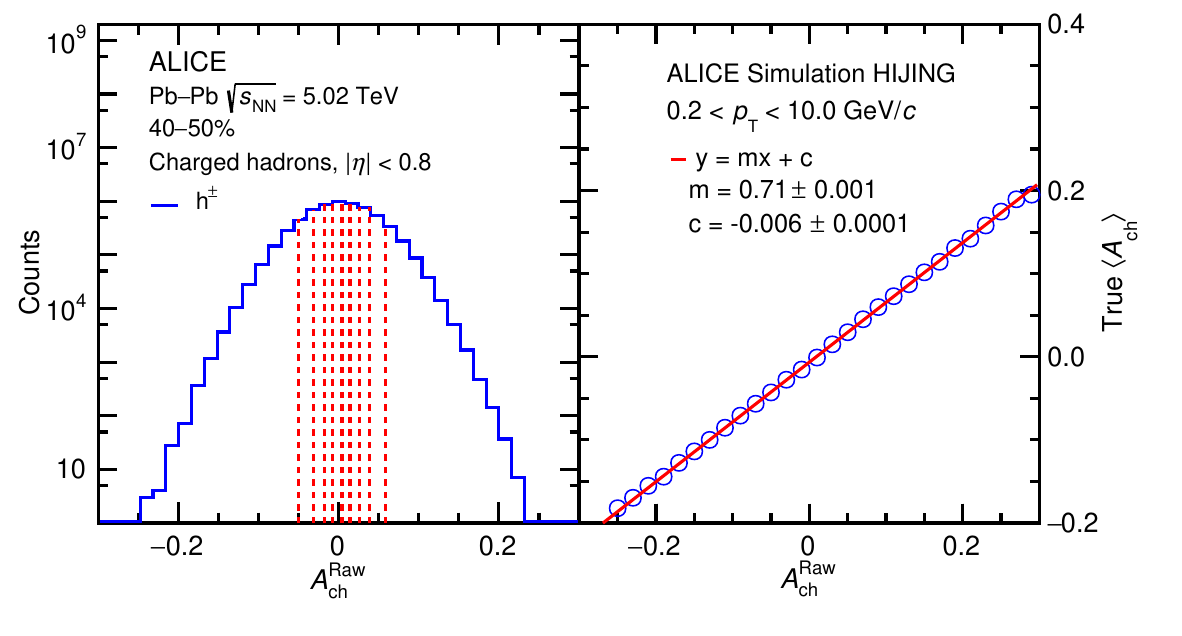}
	\caption{\label{fig:Achdist} (Left panel): Raw \Ach\xspace distribution in Pb--Pb collisions at \snn = 5.02 TeV for the 40--50\% centrality interval. Red dotted lines depict the edges of the ten \Ach\xspace classes. (Right panel): Correlation between true and raw \Ach\xspace obtained from HIJING simulations combined with a GEANT3 detector model for Pb--Pb collisions at \snn = 5.02 TeV in the 40--50\% centrality interval.} 
\end{figure}

An example of the measured raw \Ach\xspace distribution is shown in the left panel of Fig.~\ref{fig:Achdist} for the 40--50\% centrality interval. The raw \Ach\xspace distribution is divided into ten \Ach\xspace intervals, each roughly containing equal number of events. The edges of the ten \Ach\xspace classes are displayed by the dashed lines in the left panel of Fig.~\ref{fig:Achdist}. The raw \Ach\xspace is corrected to account for the limited detector acceptance and the reconstruction and identification efficiency of charged hadrons. Using simulations based on the HIJING event generator~\cite{Gyulassy:1994ew} combined with the GEANT3 model~\cite{Brun:1987ma} for particle transport in the detector material, a correlation is built between the raw and the true values of \Ach\xspace~\cite{STAR:2022hfy}. A linear fit to this correlation is performed and the fit function is used to map the raw \Ach\space to the true \Ach\xspace as shown in the right panel of Fig.~\ref{fig:Achdist}.

Within each \Ach\xspace interval, the flow coefficients \vtwo\xspace and \vthree\xspace are measured separately for positively and negatively (identified and inclusive) charged hadrons. The measurements are performed using the two-particle cumulant method~\cite{Bilandzic:2010jr} with a pseudorapidity gap of $|\Delta \eta| >$ 0.4 to suppress non-flow, i.e. correlations not related to the reaction plane~\cite{Zhou:2015iba}. The normalized slope parameters, \rnormvtwo\xspace and \rnormvthree, are then calculated for various centrality intervals with Eq.~\ref{eq:normDeltaVn} using the values of \Ach\xspace corrected for detector effects as described above.

The ESE technique is further employed to estimate possible CMW signal contribution to the $\Delta$IC as proposed in Ref.~\cite{Wang:2021nvh}. In particular, the residual magnitude of this observable when \vtwo\xspace goes to zero can be used to disentangle the potential CMW signal from the background contributions~\cite{Acharya:2017fau}. The second-order reduced flow vector $q_{2}$ is used for event shape selection as in Ref.~\cite{Acharya:2017fau}. It is defined as
\begin{equation}
	q_2= Q_2 / \sqrt{\rm M},
\end{equation}
where $Q_2$ is the magnitude of the flow vector and M is the multiplicity. The $Q_2$ is calculated from the azimuthal distribution of the energy deposited in the V0C detector. In order to account for a non-uniform detector response, the V0 detector is calibrated using two procedures: gain equalization and recentering. The former flattens the mean multiplicity distribution of the eight azimuthal sectors in each ring, while the latter corrects for systematic shifts of the mean values of the $Q_2$ vector components~\cite{Poskanzer:1998yz}. For each centrality interval, ten $q_{2}$ ranges are explored, ranging from the most elliptic to the most isotropic event classes.

To separate the LCC background contributions from the potential CMW signal, the dependence of the $\Delta$IC on \vtwo\xspace (defined in Section~\ref{Section:Introduction}) is fitted with a linear function.
The CMW fraction to the $\Delta$IC is then obtained by the ratio between the observable at zero \vtwo\xspace and at finite \vtwo
\begin{equation} \label{eq:fcmw}
	f_{\rm CMW} \equiv \frac{b}{a~\langle v_2 \rangle + b},
\end{equation}

where $a$ and $b$ are the slope and the intercept of the linear function, respectively, and $\langle{\vtwo}\rangle$ is the average value over the measured \vtwo\xspace range.

\section{Systematic uncertainties} 
\label{Section:Systematics}

To estimate the systematic uncertainties on the normalized slopes and $f_{\rm CMW}$, the event and track selection criteria are varied from their nominal values. Table~\ref{Tab:SystematicSources} provides a list of variables used in the selections along with their default and varied values. These include modifying the range of $z$ coordinate of the primary vertex ($V_{z}$) from $|V_{z}|\le$10~cm to $|V_{z}|\le$8~cm to examine the detector acceptance dependence. The impact of the track-quality selections is evaluated by changing the minimum number of TPC space points from 70 to 80 and varying the $\chi^{2}_{\rm TPC}$ per TPC space point from 2.5 to 2.0. To test the influence of the contamination from secondary particles on the measurement, tighter selection criteria than the nominal ones are applied to both DCA$_{xy}$ and DCA$_{z}$. To estimate the effects of non-flow contributions, the pseudorapidity gap ($\Delta\eta$) is varied from $|\Delta\eta|\ge$ 0.4 to $|\Delta\eta|\ge$ 0.6 for charged hadrons and pions, and to $|\Delta\eta|\ge$ 0.5 for kaons and protons. Particle identification criteria, namely \nsigmaTPC\xspace and \nsigmaTOF\xspace, are also subject to variations to account for potential systematic effects in the particle identification process and their impact on the final analysis results. The reconstruction efficiency for charged hadrons is calculated using only pions, kaons, and protons and the differences observed in the results are incorporated as systematic uncertainties. For each systematic variation, the corrections for the non-uniform acceptance and for the reconstruction efficiency of inclusive and identified charged hadrons are estimated using collision data and MC simulations. 
To identify the statistically significant systematic sources, the ratio B = Y/$\sigma_{\rm B}$ is calculated, where Y represents the difference between the results with the default and the modified selections, and $\sigma_{B}$ is the error of the difference estimated as $\sqrt{|\sigma^{2}_{\rm default} \pm \sigma^{2}_{\rm varied}|}$, where \enquote{+} is used for the uncorrelated and \enquote{-} for the correlated sources. The statistical uncertainties $\sigma_{\rm default}$ and $\sigma_{\rm varied}$ are estimated separately for the results using the default and varied event/particle selection criteria, following a subsampling method, with 20 equally sized independent samples. 
Each variation that exhibits a significant difference from the nominal result by more than 1$\sigma_{\rm B}$, according to the recommendations from Ref.~\cite{Barlow:2002yb}, is considered a source of systematic uncertainty. 
The total systematic uncertainties are then obtained by summing the different contributions in quadrature.

\begin{table}[h!]
	\caption{Nominal event and track selection criteria and the corresponding variations used for the estimation of the systematic uncertainties.}
	\begin{center}
		\begin{tabular}{ |l|c|c| }
			\hline
			(No.) Source & Default Value & Variations \\ 
			\hline
			(1)  Primary $|V_{z}|$ & $ <$ 10 cm & $<$8 cm \\  
			\hline
			(2) TPC space points & $>$70 & $>$80 \\
			\hline 
			(3) $\chi^{2}_{\mathrm{TPC}}$/cluster & $<$2.5 & $<$2.0 \\  
			\hline
			(4) DCA$_{xy}$ (DCA$_{z}$) &  $<$2.4 (3.2) cm  &  $<$7(0.0026 + (0.005/$p_{\mathrm{T}}^{1.01}$)), (2.0) cm \\ 
			\hline 
			(5) $|\Delta \eta|$ &  $>$0.4  &  $>$0.6 (0.5 for K and p) \\ 
			\hline
			(6) PID ($\pi$) & - & - \\
			0.2 $ < p_{\mathrm{T}} <$ 0.5 (GeV/$c$) & $|n\sigma|_{\rm TPC}<$ 3 & $|n\sigma|_{\rm TPC}<$ 2.5 \\
			0.5 $ < p_{\mathrm{T}} <$ 2.0 (GeV/$c$) & $\sqrt{|n\sigma|_{\rm TPC}^{2} + |n\sigma|_{\rm TOF}^{2}}<$ 3 & $\sqrt{|n\sigma|_{\rm TPC}^{2} + |n\sigma|_{\rm TOF}^{2}}<$ 3\\
			PID (K) & - & - \\
			0.2 $ < p_{\mathrm{T}} <$ 0.45 (GeV/$c$) & $|n\sigma|_{\rm TPC}<$ 3 & $|n\sigma|_{\rm TPC}<$ 2.5\\
			0.45 $ < p_{\mathrm{T}} <$ 2.0 (GeV/$c$) & $\sqrt{|n\sigma|_{\rm TPC}^{2} + |n\sigma|_{\rm TOF}^{2}}<$ 2.5 & $\sqrt{|n\sigma|_{\rm TPC}^{2} + |n\sigma|_{\rm TOF}^{2}}<$ 2\\
			PID (p) & - & - \\
			0.5 $ < p_{\mathrm{T}} <$ 0.6 (GeV/$c$) & $|n\sigma|_{\rm TPC}<$ 3 & $|n\sigma|_{\rm TPC}<$ 3.5 \\
			0.6 $ < p_{\mathrm{T}} <$ 2.0 (GeV/$c$) & $\sqrt{|n\sigma|_{\rm TPC}^{2} + |n\sigma|_{\rm TOF}^{2}}<$ 3.0 & $\sqrt{|n\sigma|_{\rm TPC}^{2} + |n\sigma|_{\rm TOF}^{2}}<$ 3.5 \\
			\hline
			(7) Efficiency calculation &  All unidentified charged hadrons  & Identified charged hadrons ($\pi$+K+p)  \\ 
			\hline
		\end{tabular}
		\label{Tab:SystematicSources}
	\end{center}
\end{table}
Table~\ref{Tab:SystematicValuesnorm} summarizes
the maximum magnitude of the systematic uncertainties on the normalized slopes over all the centrality intervals from each individual source which passes the criteria described above. 
The systematic sources for the $f_{\rm CMW}$ observable are only a few, namely Primary $V_{z}$, $\chi^{2}_{\mathrm{TPC}}$/cluster, and DCA selections. The associated uncertainties are 0.024, 0.047, and 0.068, respectively. 

\begin{table}[h]
	\small
	\caption{Maximum systematic uncertainty (absolute value) on normalized slope per particle species over all centrality intervals from individual sources (see
		Table~\ref{Tab:SystematicSources} for an explanation of
		each source).} 
	\vspace{-0.5cm}
	\begin{center}
		\begin{tabular}{|c|c|c|c|c|c|c|c|c|}
			\hline
			\multirow{3}{*}{Sources} & \multicolumn{2}{c|}{ inclusive (\hpm) } & 
			\multicolumn{2}{c|}{ \pipm} &
			\multicolumn{2}{c|}{ \Kpm} &
			\multicolumn{2}{c|}{ p$+\overline{\rm p}$} \\
			\cline{2-9}
			\cline{2-9}
			&  $r_{2}^{\mathrm{Norm}}$ & $r_{3}^{\mathrm{Norm}}$  & $r_{2}^{\mathrm{Norm}}$
			& $r_{3}^{\mathrm{Norm}}$   & $r_{2}^{\mathrm{Norm}}$ & $r_{3}^{\mathrm{Norm}}$ & $r_{2}^{\mathrm{Norm}}$ & $r_{3}^{\mathrm{Norm}}$ \\
			\hline
			(1) Primary $V_{z}$  & 0.014 & 0.03 & 0.012 & 0.03 & 0.019 & 0.12 & 0.02 & 0.021 \\ \hline
			(2) TPC space points  & 0.003 & 0.033 & 0.01 & 0.033 & 0.033 & 0.23 & 0.036 & 0.14 \\ \hline
			(3) $\chi^{2}_{\mathrm{TPC}}$/cluster  & 0.009  & 0.047 & 0.0002 & 0.047 & 0.02 & 0.31 & 0.035 & 0.18 \\ \hline
			(4)  DCA$_{xy}$ (DCA$_{z}$)  & 0.005 & 0.044 & 0.023 & 0.044 & 0.02 & 0.18 & 0.026 & 0.19 \\ \hline
			(5) $|\Delta \eta|$  & 0.013 & 0.09 & 0.018 & 0.09 & 0.017 & 0.31 & 0.052 & 0.11 \\ \hline
			(6) PID  & - & - & 0.05 & 0.05 & 0.013 & 0.11 & 0.004 & 0.13 \\ \hline
			(7) Efficiency  & 0.032 & 0.049 & - & - & - & - & - & - \\ \hline
		\end{tabular}
		\label{Tab:SystematicValuesnorm}
	\end{center}
\end{table}

\section{Results} 
\label{Section:Results}

\subsection{$A_{\rm ch}$ dependence of $v_n$ and centrality dependence of the slope $r_n^{\rm Norm}$}
\begin{figure}[!ht]
	\centering
	\includegraphics[height=0.75\textwidth]{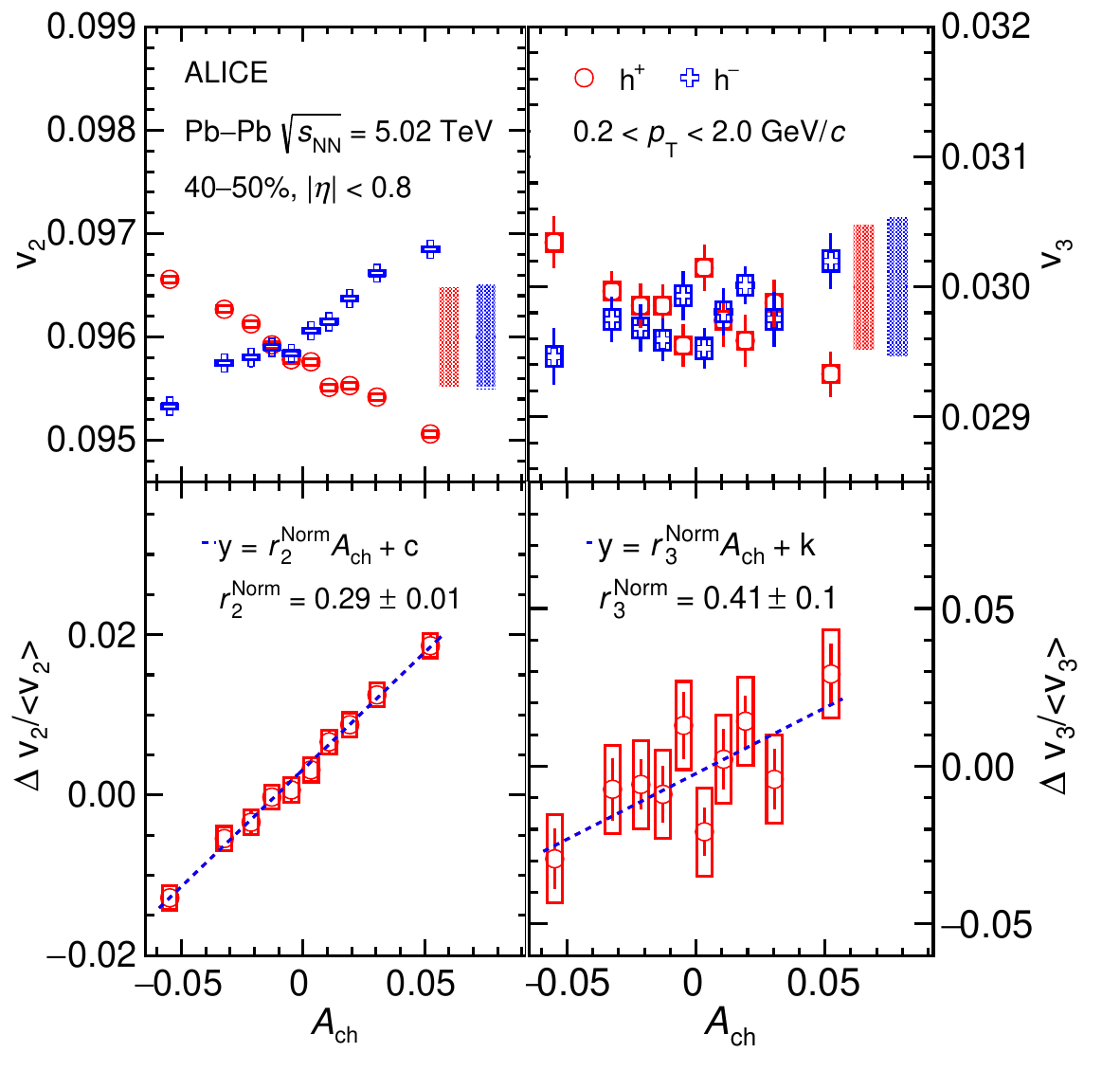}
	\caption{The top left panel shows the \vtwo\space of positively (red markers) and negatively (blue markers) charged hadrons as a function of the corrected \Ach\space, while the top right panel shows the same for \vthree. Statistical uncertainties are shown by bars and uncorrelated (correlated) systematic uncertainties by open boxes (shaded bands). The bottom left panel shows $\Delta v_{2}/\langle v_{2} \rangle $ as a function of the corrected \Ach\xspace and bottom right panel shows the same for $\Delta v_{3}/\langle v_{3} \rangle $, all for the 40--50\% centrality interval in Pb$-$Pb collisions at \snn = 5.02 TeV. The dotted blue line shows the linear fit to the data points to obtain the values of normalized slopes (\rnormvtwo\xspace and \rnormvthree).} 
	\label{fig:v2v3delv2v3}
\end{figure}
The top left and top right panels of Fig.~\ref{fig:v2v3delv2v3} show, respectively, the \vtwo\xspace and \vthree\xspace of positively and negatively charged hadrons as a function of \Ach\xspace for Pb--Pb collisions at \snn $=$ 5.02 TeV in the 40--50\% centrality interval. A significant decreasing (increasing) trend of \vtwo\xspace for positively (negatively) charged hadrons as a function of \Ach\xspace is observed. Such trends are also present for the \vthree\xspace coefficient though the fluctuations are larger. In Fig.~\ref{fig:v2v3delv2v3} the correlated uncertainties between the hadrons and among \Ach\xspace intervals are represented with shaded bands. The normalized difference between the \vtwo\xspace and \vthree\xspace of positive and negative charged particles, as defined in Eq.~\ref{eq:normDeltaVn}, is shown as a function of \Ach\xspace in the bottom left and right panels of Fig.~\ref{fig:v2v3delv2v3}, respectively. They are fitted with linear functions to obtain the slopes, \rnormvtwo\xspace and \rnormvthree. 
In addition to a positive slope for $\Delta v_{2}/\langle v_{2} \rangle $, a non-zero slope is also found for $\Delta v_{3}/\langle v_{3} \rangle $ (i.e., \rnormvthree $>0$), which mainly contains the background correlations. This non-zero \rnormvthree\ indicates that similar background is also present in \rnormvtwo\xspace which need to be subtracted to make a quantitative measurement of CMW. 
Note that the difference in non-flow contributions to same-sign and opposite-sign pairs can induce a trivial correlation between \vn\xspace and \Ach~\cite{Xu:2020sln, STAR:2022hfy}. However, it is estimated that this trivial correlation is negligible at this collision energy. 
\begin{figure}[!hbt]
	\vspace{0.5cm}
	\centering
	\includegraphics[height=0.5\textwidth]{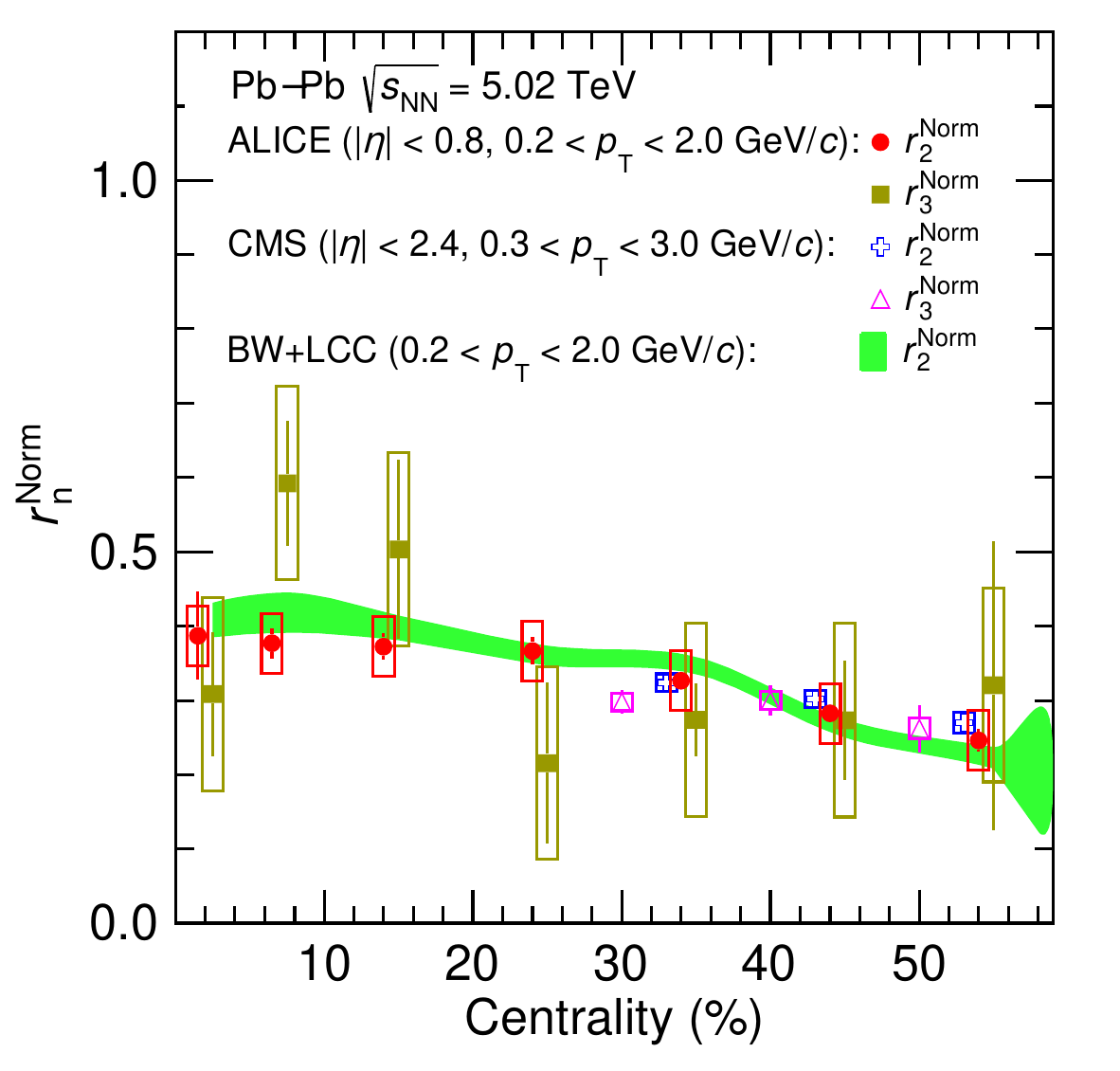}
	\caption{Centrality dependence of normalized slopes \rnormvtwo\xspace and \rnormvthree\xspace for inclusive charged hadrons in Pb$-$Pb collisions at \snn = 5.02 TeV compared with CMS results~\cite{CMS:2017pah} and a BW+LCC model calculation~\cite{wu_global_2023}. Statistical (systematic) uncertainties are depicted by bars (boxes). ALICE \rnormvtwo\xspace and \rnormvthree\space and CMS \rnormvtwo\xspace data points  are slightly shifted horizontally for visibility.} 
	\label{fig:normslopecompare}
\end{figure}
\begin{figure}[!hbt]
	\centering
	\includegraphics[height=0.55\textwidth]{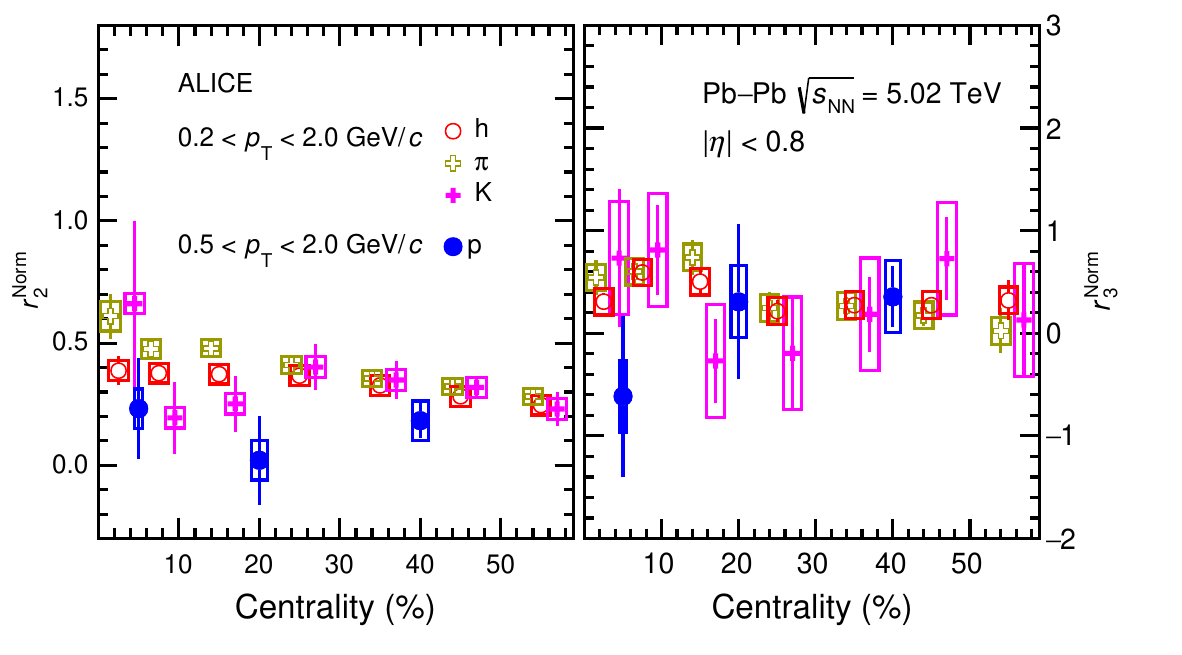}
	\caption{Centrality dependence of normalized slopes \rnormvtwo\xspace(left panel) and \rnormvthree\xspace(right panel) for inclusive and identified charged hadrons in Pb--Pb collisions at \snn = 5.02 TeV. Statistical (systematic) uncertainties are depicted by bars (boxes). The data points for charged pions and kaons are slightly shifted horizontally for visibility.}
	\label{fig:pidnormslope}
\end{figure}
Figure~\ref{fig:normslopecompare} shows \rnormvtwo\xspace (red markers) and \rnormvthree\xspace (green markers) as a function of centrality for inclusive charged hadrons in Pb--Pb collisions at \snn = 5.02 TeV. 
Despite the large uncertainties, \rnormvthree\xspace  seems to be of similar magnitude to \rnormvtwo\xspace, for most of the centrality range.
The CMW is expected to develop in the direction of the magnetic field \textbf{B}, i.e., approximately perpendicular to the second order event plane rather than the third order. The third order plane has very little correlation with the second order event plane~\cite{ALICE:2016kpq, ALICE:2011ab, ATLAS:2014ndd}. Hence, the non-zero \rnormvthree\xspace value cannot originate from the CMW-induced electric quadrupole moment, but it should rather be attributed to the LCC mechanism. Thus, the similarity between the magnitudes of \rnormvtwo\xspace and \rnormvthree\xspace could indicate that both of them are mainly dominated by the LCC background. The results are also in good agreement with the CMS measurements for the same collision system and energy~\cite{CMS:2017pah}. While the precision of the CMS measurements is notable, the findings presented in this study expand the scope of measurements to encompass the most central collision scenarios. 
The \rnormvtwo\xspace is further compared with a blast wave model calculation (green band) which takes into account the LCC effect~\cite{wu_global_2023}. This model uses blast wave parameters obtained from the simultaneous fit of \pt-differential yields and \vtwo\xspace of identified particles from Pb$-$Pb collisions at \snn $=$ 5.02 TeV~\cite{ALICE:2020siw}. Then, the model distributes source points within an ellipsoid (defined by the blast wave parameter) from where it produces oppositely charged particles (i.e. balancing charges). The number of sources is tuned in such a way that the \Ach\xspace distribution from the output of the model matches with the experimental data. Then, the normalized slope values are calculated from the model, following the same procedure as in data.
The agreement between the model and the experimental results of \rnormvtwo\xspace points to a large background contribution in the measurement.
The \rnormvtwo\xspace and \rnormvthree\xspace results for identified hadrons are shown in Fig.~\ref{fig:pidnormslope} as a function of centrality. For \rnormvtwo, the slope for kaons behaves similarly as the slope for pions, while the slope for protons slightly differs with a weak \Ach--$\Delta$\vtwo\xspace dependence on centrality. In the 20\% most central collisions, a hint of mass ordering is observed, i.e., the slopes of pions are slightly larger than those of kaons, and the slopes of kaons are slightly larger than those of protons. However, the uncertainties are too large for a definitive conclusion to be made. The LCC process, the dominant background for this measurement, can reproduce the measurement of \rnormvtwo\ in data. Therefore, it is expected that the mass ordering of $v_{2}$ of identified particles might be the reason behind the apparent particle type dependence of \rnormvtwo\ seen in Fig.~\ref{fig:pidnormslope}. This finding contradicts the predictions of the hydrodynamic study, which anticipated negative slopes for kaons and protons~\cite{Hatta:2015hca}.
A similar behavior is found in the STAR CMW measurement at a lower collision energy, in which the isospin and strangeness chemical potentials and the different absorption cross section are expected to play a role~\cite{STAR:2022hfy, Hatta:2015hca, Hatta:2016czn}. 
However, at LHC energies, the values of the chemical potentials are close to zero, suggesting that their influence should be negligible. For \rnormvthree, the slopes of all the measured hadron species are compatible with each other within the uncertainties. Since no predictions for the CMW for different particle species are available at LHC energies, the results shown here provide, for the first time, an experimental input for theoretical calculations.

\begin{figure}[!hbt]
	\centering
	\begin{subfigure}
		\centering
		\includegraphics[height=0.68\textwidth]{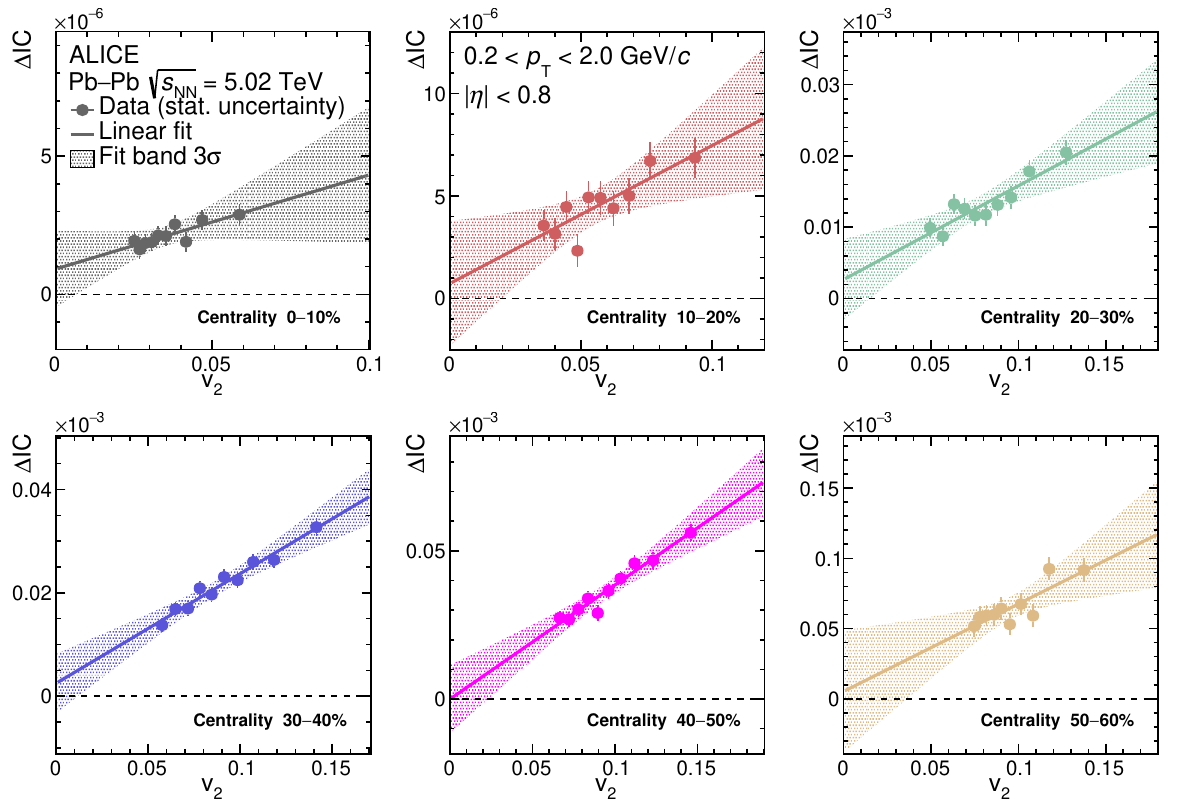}
	\end{subfigure}
	\caption{Dependence of $\Delta\rm{IC}$ on \vtwo\xspace of shape-selected events from the 0--10\% (top left panel) to the 50--60\% (bottom right panel) centrality intervals of Pb--Pb collisions at \snn = 5.02 TeV. The solid lines are straight line fit to the data points. Only statistical uncertainties are shown. The bands represent the three standard deviation uncertainties from the linear fit.}
	\label{fig:ese}
\end{figure}
\subsection{Constraining the fraction of the CMW with the ESE}

Figure~\ref{fig:ese} shows the difference of the integrated covariance $\Delta$IC as a function of \vtwo\xspace for $q_{2}$-selected events in six centrality intervals. Solid lines, representing linear fits, and colored bands, denoting three standard deviation uncertainties of the linear fits are also shown. The $\Delta$IC values exhibit a decrease as \vtwo\xspace approaches zero, as discussed in Ref.~\cite{Wang:2021nvh}. The small magnitude of the intercepts in all centrality intervals indicates that the measurement is dominated by the LCC background.

\begin{figure}[!hbt]
	\centering
	\includegraphics[height=0.5\textwidth]{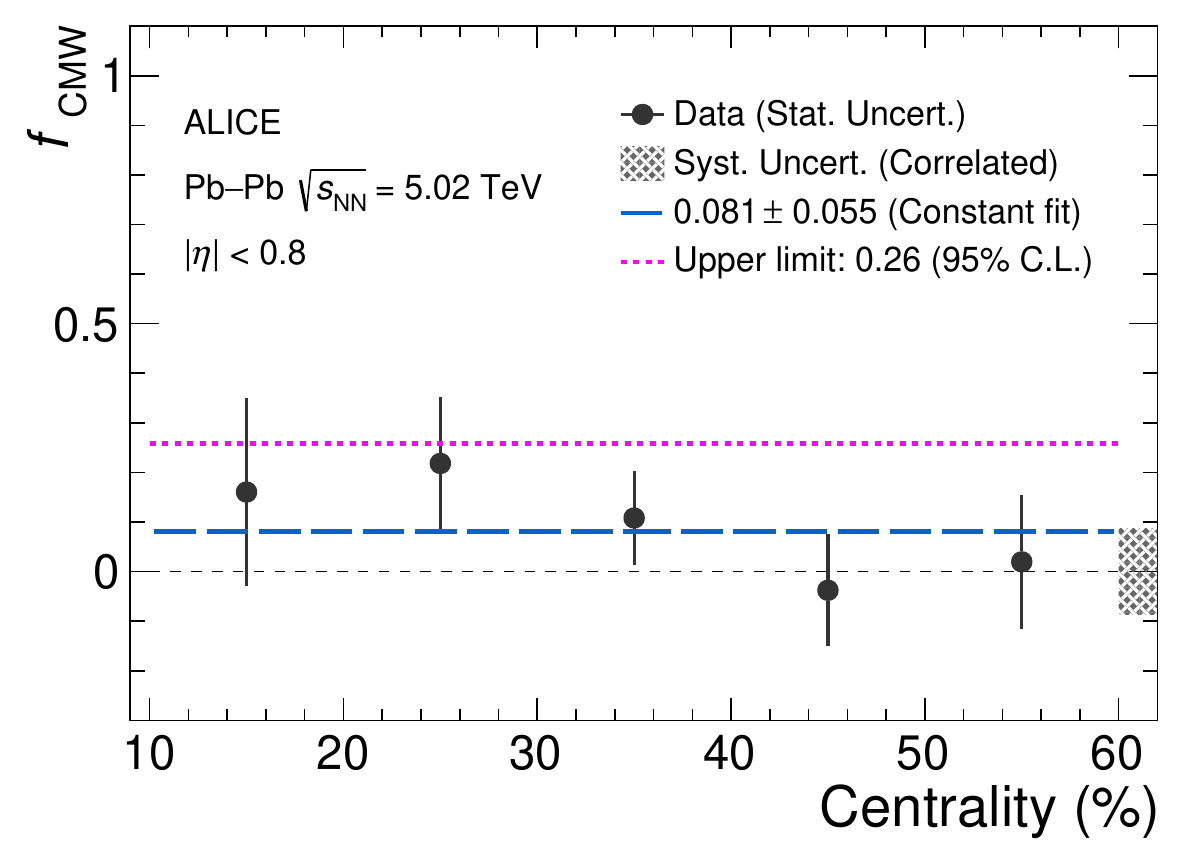}
	\caption{Centrality dependence of the extracted CMW fraction. The 95\% confidence level of the upper limit is also shown by the dotted magenta line. Statistical uncertainties are depicted by bars, while the correlated systematic uncertainty is represented by a shaded band on the right edge. The blue line is the constant fit line of the data points.}
	\label{fig:fcmw} 
\end{figure}

After obtaining the slope and intercept from the fit of $\Delta$IC as a function of \vtwo, the fraction of the CMW signal, denoted as $f_{\rm CMW}$, can be determined using Eq.~\ref{eq:fcmw}. The centrality dependence of $f_{\rm CMW}$ is presented in Fig.~\ref{fig:fcmw}\xspace where the error bars display the statistical uncertainties extracted from the fits of Fig.~\ref{fig:ese}. 
Meaningful measurements are not possible in the most central and peripheral collisions, due to the small \vtwo\xspace values and the large statistical fluctuations. Therefore, the $f_{\rm CMW}$ results are reported only in the 10--60\% centrality range. The systematic uncertainties are estimated for the different possible sources, as discussed in Sec.~\ref{Section:Systematics}.
The systematic sources, which are found to be significant among centrality intervals, are combined in quadrature and shown as a dark shaded band in Fig.~\ref{fig:fcmw}, at centrality around 60\%. 
It is found that the $f_{\rm CMW}$ is consistent with zero within the reported statistical and systematic uncertainties.

The $f_{\rm CMW}$ data points in Fig.~\ref{fig:fcmw} are fitted with a constant, yielding a value of $f_{\rm CMW} = 0.081 \pm 0.055$(the dashed blue line), which should be combined with the systematic uncertainty of 0.087(the gray box on the edge) to establish an upper limit on $f_{\rm CMW}$ of 26\% (38\%) at 95\% (99.7\%) confidence level shown by the dotted magenta line. 
This result provides the first quantitative assessment of the upper limit of the fraction of the Chiral Magnetic Wave (CMW) at the top LHC energy.

\section{Summary} 
\label{Section:Summary}

The difference between the \vtwo\xspace and \vthree\xspace coefficients of positively and negatively charged particles are measured as a function of the charge asymmetry \Ach\xspace for inclusive and identified hadrons in Pb--Pb collisions at \snn $=$ 5.02 TeV. The slopes \rnormvtwo\xspace and \rnormvthree\xspace are found to be consistent with each other within the reported uncertainties, suggesting that the dominant contribution to \rnormvtwo\xspace is not due to CMW. A blast wave parameterization that incorporates local charge conservation, tuned to reproduce the \Ach\xspace distribution, is able to fully describe the magnitude of \rnormvtwo\xspace which is expected to be sensitive to CMW. Additionally, a hint for mass ordering of the \rnormvtwo\xspace slopes of pions, kaons, and protons is discernible in the most central collisions, though it is accompanied by large uncertainties.
Furthermore, using the Event Shape Engineering (ESE) technique, both the fraction and the upper limit of the CMW signal are extracted. Averaging over the 10--60\% centrality interval, the CMW fraction is consistent with zero within uncertainties and an upper limit of 26\% (38\%) is estimated at 95\% (99.7\%) confidence level.


\newenvironment{acknowledgement}{\relax}{\relax}
\begin{acknowledgement}
\section*{Acknowledgements}

The ALICE Collaboration would like to thank all its engineers and technicians for their invaluable contributions to the construction of the experiment and the CERN accelerator teams for the outstanding performance of the LHC complex.
The ALICE Collaboration gratefully acknowledges the resources and support provided by all Grid centres and the Worldwide LHC Computing Grid (WLCG) collaboration.
The ALICE Collaboration acknowledges the following funding agencies for their support in building and running the ALICE detector:
A. I. Alikhanyan National Science Laboratory (Yerevan Physics Institute) Foundation (ANSL), State Committee of Science and World Federation of Scientists (WFS), Armenia;
Austrian Academy of Sciences, Austrian Science Fund (FWF): [M 2467-N36] and Nationalstiftung f\"{u}r Forschung, Technologie und Entwicklung, Austria;
Ministry of Communications and High Technologies, National Nuclear Research Center, Azerbaijan;
Conselho Nacional de Desenvolvimento Cient\'{\i}fico e Tecnol\'{o}gico (CNPq), Financiadora de Estudos e Projetos (Finep), Funda\c{c}\~{a}o de Amparo \`{a} Pesquisa do Estado de S\~{a}o Paulo (FAPESP) and Universidade Federal do Rio Grande do Sul (UFRGS), Brazil;
Bulgarian Ministry of Education and Science, within the National Roadmap for Research Infrastructures 2020--2027 (object CERN), Bulgaria;
Ministry of Education of China (MOEC) , Ministry of Science \& Technology of China (MSTC) and National Natural Science Foundation of China (NSFC), China;
Ministry of Science and Education and Croatian Science Foundation, Croatia;
Centro de Aplicaciones Tecnol\'{o}gicas y Desarrollo Nuclear (CEADEN), Cubaenerg\'{\i}a, Cuba;
Ministry of Education, Youth and Sports of the Czech Republic, Czech Republic;
The Danish Council for Independent Research | Natural Sciences, the VILLUM FONDEN and Danish National Research Foundation (DNRF), Denmark;
Helsinki Institute of Physics (HIP), Finland;
Commissariat \`{a} l'Energie Atomique (CEA) and Institut National de Physique Nucl\'{e}aire et de Physique des Particules (IN2P3) and Centre National de la Recherche Scientifique (CNRS), France;
Bundesministerium f\"{u}r Bildung und Forschung (BMBF) and GSI Helmholtzzentrum f\"{u}r Schwerionenforschung GmbH, Germany;
General Secretariat for Research and Technology, Ministry of Education, Research and Religions, Greece;
National Research, Development and Innovation Office, Hungary;
Department of Atomic Energy Government of India (DAE), Department of Science and Technology, Government of India (DST), University Grants Commission, Government of India (UGC) and Council of Scientific and Industrial Research (CSIR), India;
National Research and Innovation Agency - BRIN, Indonesia;
Istituto Nazionale di Fisica Nucleare (INFN), Italy;
Japanese Ministry of Education, Culture, Sports, Science and Technology (MEXT) and Japan Society for the Promotion of Science (JSPS) KAKENHI, Japan;
Consejo Nacional de Ciencia (CONACYT) y Tecnolog\'{i}a, through Fondo de Cooperaci\'{o}n Internacional en Ciencia y Tecnolog\'{i}a (FONCICYT) and Direcci\'{o}n General de Asuntos del Personal Academico (DGAPA), Mexico;
Nederlandse Organisatie voor Wetenschappelijk Onderzoek (NWO), Netherlands;
The Research Council of Norway, Norway;
Commission on Science and Technology for Sustainable Development in the South (COMSATS), Pakistan;
Pontificia Universidad Cat\'{o}lica del Per\'{u}, Peru;
Ministry of Education and Science, National Science Centre and WUT ID-UB, Poland;
Korea Institute of Science and Technology Information and National Research Foundation of Korea (NRF), Republic of Korea;
Ministry of Education and Scientific Research, Institute of Atomic Physics, Ministry of Research and Innovation and Institute of Atomic Physics and University Politehnica of Bucharest, Romania;
Ministry of Education, Science, Research and Sport of the Slovak Republic, Slovakia;
National Research Foundation of South Africa, South Africa;
Swedish Research Council (VR) and Knut \& Alice Wallenberg Foundation (KAW), Sweden;
European Organization for Nuclear Research, Switzerland;
Suranaree University of Technology (SUT), National Science and Technology Development Agency (NSTDA), Thailand Science Research and Innovation (TSRI) and National Science, Research and Innovation Fund (NSRF), Thailand;
Turkish Energy, Nuclear and Mineral Research Agency (TENMAK), Turkey;
National Academy of  Sciences of Ukraine, Ukraine;
Science and Technology Facilities Council (STFC), United Kingdom;
National Science Foundation of the United States of America (NSF) and United States Department of Energy, Office of Nuclear Physics (DOE NP), United States of America.
In addition, individual groups or members have received support from:
European Research Council, Strong 2020 - Horizon 2020 (grant nos. 950692, 824093), European Union;
Academy of Finland (Center of Excellence in Quark Matter) (grant nos. 346327, 346328), Finland.

\end{acknowledgement}

\bibliographystyle{utphys}   
\bibliography{bibliography}

\providecommand{\href}[2]{#2}\begingroup\raggedright\begin{thebibliography}{10}

\bibitem{Busza:2018rrf}
W.~Busza, K.~Rajagopal, and W.~van~der Schee, ``{Heavy Ion Collisions: The Big
  Picture, and the Big Questions}'',
  \href{http://dx.doi.org/10.1146/annurev-nucl-101917-020852}{{\em Ann. Rev.
  Nucl. Part. Sci.} {\bfseries 68} (2018) 339--376},
  \href{http://arxiv.org/abs/1802.04801}{{\ttfamily arXiv:1802.04801
  [hep-ph]}}.

\bibitem{Schukraft:1993fcq}
J.~Schukraft, ``{Ultra-relativistic heavy-ion collisions: Searching for the
  quark-gluon plasma}'',
  \href{http://dx.doi.org/10.1016/0375-9474(93)90613-3}{{\em Nucl. Phys. A}
  {\bfseries 553} (1993) 31--44}.

\bibitem{Heinz:2008tv}
U.~W. Heinz, ``{The Strongly coupled quark-gluon plasma created at RHIC}'',
  \href{http://dx.doi.org/10.1088/1751-8113/42/21/214003}{{\em J. Phys. A}
  {\bfseries 42} (2009) 214003},
  \href{http://arxiv.org/abs/0810.5529}{{\ttfamily arXiv:0810.5529 [nucl-th]}}.

\bibitem{Shuryak:1980tp}
E.~V. Shuryak, ``{Quantum Chromodynamics and the Theory of Superdense
  Matter}'', \href{http://dx.doi.org/10.1016/0370-1573(80)90105-2}{{\em Phys.
  Rept.} {\bfseries 61} (1980) 71--158}.

\bibitem{Collins:1974ky}
J.~C. Collins and M.~J. Perry, ``{Superdense Matter: Neutrons Or Asymptotically
  Free Quarks?}'', \href{http://dx.doi.org/10.1103/PhysRevLett.34.1353}{{\em
  Phys. Rev. Lett.} {\bfseries 34} (1975) 1353}.

\bibitem{ALICE:2022wpn}
{\bfseries ALICE} Collaboration, ``{The ALICE experiment -- A journey through
  QCD}'', \href{http://arxiv.org/abs/2211.04384}{{\ttfamily arXiv:2211.04384
  [nucl-ex]}}.

\bibitem{Aoki:2006we}
Y.~Aoki, G.~Endrodi, Z.~Fodor, S.~D. Katz, and K.~K. Szabo, ``{The Order of the
  quantum chromodynamics transition predicted by the standard model of particle
  physics}'', \href{http://dx.doi.org/10.1038/nature05120}{{\em Nature}
  {\bfseries 443} (2006) 675--678},
  \href{http://arxiv.org/abs/hep-lat/0611014}{{\ttfamily
  arXiv:hep-lat/0611014}}.

\bibitem{Brown:1990ev}
F.~R. Brown~et al, ``{On the existence of a phase transition for QCD with three
  light quarks}'', \href{http://dx.doi.org/10.1103/PhysRevLett.65.2491}{{\em
  Phys. Rev. Lett.} {\bfseries 65} (1990) 2491--2494}.

\bibitem{Bzdak:2011yy}
A.~Bzdak and V.~Skokov, ``{Event-by-event fluctuations of magnetic and electric
  fields in heavy ion collisions}'',
  \href{http://dx.doi.org/10.1016/j.physletb.2012.02.065}{{\em Phys. Lett.}
  {\bfseries B710} (2012) 171--174},
\href{http://arxiv.org/abs/1111.1949}{{\ttfamily arXiv:1111.1949 [hep-ph]}}.

\bibitem{Lee:1973iz}
T.~D. Lee, ``{A Theory of Spontaneous T Violation}'',
  \href{http://dx.doi.org/10.1103/PhysRevD.8.1226}{{\em Phys. Rev. D}
  {\bfseries 8} (1973) 1226--1239}.

\bibitem{Lee:1974ma}
T.~D. Lee and G.~C. Wick, ``{Vacuum Stability and Vacuum Excitation in a Spin 0
  Field Theory}'',
\href{http://dx.doi.org/10.1103/PhysRevD.9.2291}{{\em Phys. Rev.} {\bfseries
  D9} (1974) 2291--2316}.

\bibitem{Kharzeev:1998kz}
D.~Kharzeev, R.~D. Pisarski, and M.~H.~G. Tytgat, ``{Possibility of spontaneous
  parity violation in hot QCD}'',
  \href{http://dx.doi.org/10.1103/PhysRevLett.81.512}{{\em Phys. Rev. Lett.}
  {\bfseries 81} (1998) 512--515},
\href{http://arxiv.org/abs/hep-ph/9804221}{{\ttfamily arXiv:hep-ph/9804221
  [hep-ph]}}.

\bibitem{Morley:1983wr}
P.~D. Morley and I.~A. Schmidt, ``{Strong P, {CP}, T Violations in Heavy Ion
  Collisions}'',
\href{http://dx.doi.org/10.1007/BF01551807}{{\em Z. Phys.} {\bfseries C26}
  (1985) 627}.

\bibitem{Kharzeev:1999cz}
D.~Kharzeev and R.~D. Pisarski, ``{Pionic measures of parity and CP violation
  in high-energy nuclear collisions}'',
  \href{http://dx.doi.org/10.1103/PhysRevD.61.111901}{{\em Phys. Rev.}
  {\bfseries D61} (2000) 111901},
\href{http://arxiv.org/abs/hep-ph/9906401}{{\ttfamily arXiv:hep-ph/9906401
  [hep-ph]}}.

\bibitem{Kharzeev:2013ffa}
D.~E. Kharzeev, ``{The Chiral Magnetic Effect and Anomaly-Induced Transport}'',
  \href{http://dx.doi.org/10.1016/j.ppnp.2014.01.002}{{\em Prog. Part. Nucl.
  Phys.} {\bfseries 75} (2014) 133--151},
\href{http://arxiv.org/abs/1312.3348}{{\ttfamily arXiv:1312.3348 [hep-ph]}}.

\bibitem{Kharzeev:2015kna}
D.~E. Kharzeev, ``{Topology, magnetic field, and strongly interacting
  matter}'', \href{http://dx.doi.org/10.1146/annurev-nucl-102313-025420}{{\em
  Ann. Rev. Nucl. Part. Sci.} {\bfseries 65} (2015) 193--214},
\href{http://arxiv.org/abs/1501.01336}{{\ttfamily arXiv:1501.01336 [hep-ph]}}.

\bibitem{kharzeev_chiral_2021}
D.~E. Kharzeev and J.~Liao, ``Chiral magnetic effect reveals the topology of
  gauge fields in heavy-ion collisions'',
  \href{http://dx.doi.org/10.1038/s42254-020-00254-6}{{\em Nat Rev Phys}
  {\bfseries 3} (Jan., 2021) 55--63}.
  \url{http://www.nature.com/articles/s42254-020-00254-6}.

\bibitem{Voloshin:2004vk}
S.~A. Voloshin, ``{Parity violation in hot QCD: How to detect it}'',
  \href{http://dx.doi.org/10.1103/PhysRevC.70.057901}{{\em Phys. Rev.}
  {\bfseries C70} (2004) 057901},
\href{http://arxiv.org/abs/hep-ph/0406311}{{\ttfamily arXiv:hep-ph/0406311
  [hep-ph]}}.

\bibitem{Kharzeev:2007tn}
D.~Kharzeev and A.~Zhitnitsky, ``{Charge separation induced by P-odd bubbles in
  QCD matter}'', \href{http://dx.doi.org/10.1016/j.nuclphysa.2007.10.001}{{\em
  Nucl. Phys.} {\bfseries A797} (2007) 67--79},
\href{http://arxiv.org/abs/0706.1026}{{\ttfamily arXiv:0706.1026 [hep-ph]}}.

\bibitem{Fukushima:2008xe}
K.~Fukushima, D.~E. Kharzeev, and H.~J. Warringa, ``{The Chiral Magnetic
  Effect}'', \href{http://dx.doi.org/10.1103/PhysRevD.78.074033}{{\em Phys.
  Rev.} {\bfseries D78} (2008) 074033},
\href{http://arxiv.org/abs/0808.3382}{{\ttfamily arXiv:0808.3382 [hep-ph]}}.

\bibitem{Kharzeev:2007jp}
D.~E. Kharzeev, L.~D. McLerran, and H.~J. Warringa, ``{The Effects of
  topological charge change in heavy ion collisions: 'Event by event P and CP
  violation'}'', \href{http://dx.doi.org/10.1016/j.nuclphysa.2008.02.298}{{\em
  Nucl. Phys.} {\bfseries A803} (2008) 227--253},
\href{http://arxiv.org/abs/0711.0950}{{\ttfamily arXiv:0711.0950 [hep-ph]}}.

\bibitem{Li:2014bha}
Q.~Li~et al, ``{Observation of the chiral magnetic effect in ZrTe5}'',
  \href{http://dx.doi.org/10.1038/nphys3648}{{\em Nature Phys.} {\bfseries 12}
  (2016) 550--554}, \href{http://arxiv.org/abs/1412.6543}{{\ttfamily
  arXiv:1412.6543 [cond-mat.str-el]}}.

\bibitem{Liu:2020ymh}
Y.-C. Liu and X.-G. Huang, ``{Anomalous chiral transports and spin polarization
  in heavy-ion collisions}'',
  \href{http://dx.doi.org/10.1007/s41365-020-00764-z}{{\em Nucl. Sci. Tech.}
  {\bfseries 31} (2020) 56}, \href{http://arxiv.org/abs/2003.12482}{{\ttfamily
  arXiv:2003.12482 [nucl-th]}}.

\bibitem{Gao:2020vbh}
J.-H. Gao, G.-L. Ma, S.~Pu, and Q.~Wang, ``{Recent developments in chiral and
  spin polarization effects in heavy-ion collisions}'',
  \href{http://dx.doi.org/10.1007/s41365-020-00801-x}{{\em Nucl. Sci. Tech.}
  {\bfseries 31} (2020) 90}, \href{http://arxiv.org/abs/2005.10432}{{\ttfamily
  arXiv:2005.10432 [hep-ph]}}.

\bibitem{STAR:2009wot}
{\bfseries STAR} Collaboration, B.~I. Abelev {\em et~al.}, ``{Azimuthal
  Charged-Particle Correlations and Possible Local Strong Parity Violation}'',
  \href{http://dx.doi.org/10.1103/PhysRevLett.103.251601}{{\em Phys. Rev.
  Lett.} {\bfseries 103} (2009) 251601},
  \href{http://arxiv.org/abs/0909.1739}{{\ttfamily arXiv:0909.1739 [nucl-ex]}}.

\bibitem{STAR:2009tro}
{\bfseries STAR} Collaboration, B.~I. Abelev {\em et~al.}, ``{Observation of
  charge-dependent azimuthal correlations and possible local strong parity
  violation in heavy ion collisions}'',
  \href{http://dx.doi.org/10.1103/PhysRevC.81.054908}{{\em Phys. Rev. C}
  {\bfseries 81} (2010) 054908},
  \href{http://arxiv.org/abs/0909.1717}{{\ttfamily arXiv:0909.1717 [nucl-ex]}}.

\bibitem{ALICE:2012nhw}
{\bfseries ALICE} Collaboration, B.~Abelev {\em et~al.}, ``{Charge separation
  relative to the reaction plane in Pb--Pb collisions at $\sqrt{s_{\rm {NN}}}=
  2.76$ TeV}'', \href{http://dx.doi.org/10.1103/PhysRevLett.110.012301}{{\em
  Phys. Rev. Lett.} {\bfseries 110} (2013) 012301},
  \href{http://arxiv.org/abs/1207.0900}{{\ttfamily arXiv:1207.0900 [nucl-ex]}}.

\bibitem{Schlichting:2010qia}
S.~Schlichting and S.~Pratt, ``{Charge conservation at energies available at
  the BNL Relativistic Heavy Ion Collider and contributions to local parity
  violation observables}'',
  \href{http://dx.doi.org/10.1103/PhysRevC.83.014913}{{\em Phys. Rev.}
  {\bfseries C83} (2011) 014913},
\href{http://arxiv.org/abs/1009.4283}{{\ttfamily arXiv:1009.4283 [nucl-th]}}.

\bibitem{Pratt:2010zn}
S.~Pratt, S.~Schlichting, and S.~Gavin, ``{Effects of Momentum Conservation and
  Flow on Angular Correlations at RHIC}'',
  \href{http://dx.doi.org/10.1103/PhysRevC.84.024909}{{\em Phys. Rev.}
  {\bfseries C84} (2011) 024909},
\href{http://arxiv.org/abs/1011.6053}{{\ttfamily arXiv:1011.6053 [nucl-th]}}.

\bibitem{Adamczyk:2013kcb}
{\bfseries STAR} Collaboration, L.~Adamczyk {\em et~al.}, ``{Measurement of
  charge multiplicity asymmetry correlations in high-energy nucleus-nucleus
  collisions at $\sqrt{{s}_{NN}} =$ 200 GeV}'',
  \href{http://dx.doi.org/10.1103/PhysRevC.89.044908}{{\em Phys. Rev.}
  {\bfseries C89} (2014) 044908},
\href{http://arxiv.org/abs/1303.0901}{{\ttfamily arXiv:1303.0901 [nucl-ex]}}.

\bibitem{Adamczyk:2013hsi}
{\bfseries STAR} Collaboration, L.~Adamczyk {\em et~al.}, ``{Fluctuations of
  charge separation perpendicular to the event plane and local parity violation
  in $\sqrt{s_{NN}}=200$ GeV Au+Au collisions at the BNL Relativistic Heavy Ion
  Collider}'', \href{http://dx.doi.org/10.1103/PhysRevC.88.064911}{{\em Phys.
  Rev.} {\bfseries C88} (2013) 064911},
\href{http://arxiv.org/abs/1302.3802}{{\ttfamily arXiv:1302.3802 [nucl-ex]}}.

\bibitem{Adamczyk:2014mzf}
{\bfseries STAR} Collaboration, L.~Adamczyk {\em et~al.}, ``{Beam-energy
  dependence of charge separation along the magnetic field in Au+Au collisions
  at RHIC}'', \href{http://dx.doi.org/10.1103/PhysRevLett.113.052302}{{\em
  Phys. Rev. Lett.} {\bfseries 113} (2014) 052302},
\href{http://arxiv.org/abs/1404.1433}{{\ttfamily arXiv:1404.1433 [nucl-ex]}}.

\bibitem{Acharya:2017fau}
{\bfseries ALICE} Collaboration, S.~Acharya {\em et~al.}, ``{Constraining the
  magnitude of the Chiral Magnetic Effect with Event Shape Engineering in
  Pb--Pb collisions at $\sqrt{s_\mathrm{NN}}$ = 2.76 TeV}'',
  \href{http://dx.doi.org/10.1016/j.physletb.2017.12.021}{{\em Phys. Lett.}
  {\bfseries B777} (2018) 151--162},
\href{http://arxiv.org/abs/1709.04723}{{\ttfamily arXiv:1709.04723 [nucl-ex]}}.

\bibitem{CMS:2016wfo}
{\bfseries CMS} Collaboration, V.~Khachatryan {\em et~al.}, ``{Observation of
  charge-dependent azimuthal correlations in $p$-Pb collisions and its
  implication for the search for the chiral magnetic effect}'',
  \href{http://dx.doi.org/10.1103/PhysRevLett.118.122301}{{\em Phys. Rev.
  Lett.} {\bfseries 118} (2017) 122301},
  \href{http://arxiv.org/abs/1610.00263}{{\ttfamily arXiv:1610.00263
  [nucl-ex]}}.

\bibitem{STAR:2019xzd}
{\bfseries STAR} Collaboration, J.~Adam {\em et~al.}, ``{Charge-dependent pair
  correlations relative to a third particle in $p$ + Au and $d$+ Au collisions
  at RHIC}'', \href{http://dx.doi.org/10.1016/j.physletb.2019.134975}{{\em
  Phys. Lett.} {\bfseries B798} (2019) 134975},
\href{http://arxiv.org/abs/1906.03373}{{\ttfamily arXiv:1906.03373 [nucl-ex]}}.

\bibitem{Sirunyan:2017quh}
{\bfseries CMS} Collaboration, A.~M. Sirunyan {\em et~al.}, ``{Constraints on
  the chiral magnetic effect using charge-dependent azimuthal correlations in
  $p\mathrm{Pb}$ and PbPb collisions at the CERN Large Hadron Collider}'',
  \href{http://dx.doi.org/10.1103/PhysRevC.97.044912}{{\em Phys. Rev.}
  {\bfseries C97} (2018) 044912},
\href{http://arxiv.org/abs/1708.01602}{{\ttfamily arXiv:1708.01602 [nucl-ex]}}.

\bibitem{li_chiral_2020}
W.~Li and G.~Wang, ``Chiral {Magnetic} {Effects} in {Nuclear} {Collisions}'',
  \href{http://dx.doi.org/10.1146/annurev-nucl-030220-065203}{{\em Annu. Rev.
  Nucl. Part. Sci.} {\bfseries 70} (Oct., 2020) 293--321}.
  \url{https://www.annualreviews.org/doi/10.1146/annurev-nucl-030220-065203}.

\bibitem{STAR:2019bjg}
{\bfseries STAR} Collaboration, J.~Adam {\em et~al.}, ``{Methods for a blind
  analysis of isobar data collected by the STAR collaboration}'',
  \href{http://dx.doi.org/10.1007/s41365-021-00878-y}{{\em Nucl. Sci. Tech.}
  {\bfseries 32} (2021) 48}, \href{http://arxiv.org/abs/1911.00596}{{\ttfamily
  arXiv:1911.00596 [nucl-ex]}}.

\bibitem{abdallah_search_2022}
{\bfseries STAR} Collaboration, M.~S. Abdallah {\em et~al.}, ``Search for the
  chiral magnetic effect with isobar collisions at $\sqrt{s_{\rm NN}}$ = 200
  {GeV} by the {STAR} {Collaboration} at the {BNL} {Relativistic} {Heavy} {Ion}
  {Collider}'', \href{http://dx.doi.org/10.1103/PhysRevC.105.014901}{{\em Phys.
  Rev. C} {\bfseries 105} (Jan., 2022) 014901}.
  \url{https://link.aps.org/doi/10.1103/PhysRevC.105.014901}.

\bibitem{Son:2004tq}
D.~T. Son and A.~R. Zhitnitsky, ``{Quantum anomalies in dense matter}'',
  \href{http://dx.doi.org/10.1103/PhysRevD.70.074018}{{\em Phys. Rev. D}
  {\bfseries 70} (2004) 074018},
  \href{http://arxiv.org/abs/hep-ph/0405216}{{\ttfamily arXiv:hep-ph/0405216}}.

\bibitem{Metlitski:2005pr}
M.~A. Metlitski and A.~R. Zhitnitsky, ``{Anomalous axion interactions and
  topological currents in dense matter}'',
  \href{http://dx.doi.org/10.1103/PhysRevD.72.045011}{{\em Phys. Rev. D}
  {\bfseries 72} (2005) 045011},
  \href{http://arxiv.org/abs/hep-ph/0505072}{{\ttfamily arXiv:hep-ph/0505072}}.

\bibitem{Burnier:2011bf}
Y.~Burnier, D.~E. Kharzeev, J.~Liao, and H.-U. Yee, ``{Chiral magnetic wave at
  finite baryon density and the electric quadrupole moment of quark-gluon
  plasma in heavy ion collisions}'',
  \href{http://dx.doi.org/10.1103/PhysRevLett.107.052303}{{\em Phys. Rev.
  Lett.} {\bfseries 107} (2011) 052303},
  \href{http://arxiv.org/abs/1103.1307}{{\ttfamily arXiv:1103.1307 [hep-ph]}}.

\bibitem{Burnier:2012ae}
Y.~Burnier, D.~E. Kharzeev, J.~Liao, and H.~U. Yee, ``{From the chiral magnetic
  wave to the charge dependence of elliptic flow}'',
  \href{http://arxiv.org/abs/1208.2537}{{\ttfamily arXiv:1208.2537 [hep-ph]}}.

\bibitem{Kharzeev:2010gd}
D.~E. Kharzeev and H.-U. Yee, ``{Chiral Magnetic Wave}'',
  \href{http://dx.doi.org/10.1103/PhysRevD.83.085007}{{\em Phys. Rev. D}
  {\bfseries 83} (2011) 085007},
  \href{http://arxiv.org/abs/1012.6026}{{\ttfamily arXiv:1012.6026 [hep-th]}}.

\bibitem{Yee:2013cya}
H.-U. Yee and Y.~Yin, ``{Realistic Implementation of Chiral Magnetic Wave in
  Heavy Ion Collisions}'',
  \href{http://dx.doi.org/10.1103/PhysRevC.89.044909}{{\em Phys. Rev. C}
  {\bfseries 89} (2014) 044909},
  \href{http://arxiv.org/abs/1311.2574}{{\ttfamily arXiv:1311.2574 [nucl-th]}}.

\bibitem{Taghavi:2013ena}
S.~F. Taghavi and U.~A. Wiedemann, ``{Chiral magnetic wave in an expanding QCD
  fluid}'', \href{http://dx.doi.org/10.1103/PhysRevC.91.024902}{{\em Phys. Rev.
  C} {\bfseries 91} (2015) 024902},
  \href{http://arxiv.org/abs/1310.0193}{{\ttfamily arXiv:1310.0193 [hep-ph]}}.

\bibitem{Voloshin:1994mz}
S.~Voloshin and Y.~Zhang, ``{Flow study in relativistic nuclear collisions by
  Fourier expansion of Azimuthal particle distributions}'',
  \href{http://dx.doi.org/10.1007/s002880050141}{{\em Z. Phys.} {\bfseries C70}
  (1996) 665--672},
\href{http://arxiv.org/abs/hep-ph/9407282}{{\ttfamily arXiv:hep-ph/9407282
  [hep-ph]}}.

\bibitem{Poskanzer:1998yz}
A.~M. Poskanzer and S.~A. Voloshin, ``{Methods for analyzing anisotropic flow
  in relativistic nuclear collisions}'',
  \href{http://dx.doi.org/10.1103/PhysRevC.58.1671}{{\em Phys. Rev.} {\bfseries
  C58} (1998) 1671--1678},
\href{http://arxiv.org/abs/nucl-ex/9805001}{{\ttfamily arXiv:nucl-ex/9805001
  [nucl-ex]}}.

\bibitem{wang_number--constituent-quark_2022}
M.~Wang, J.-Q. Tao, H.~Zheng, W.-C. Zhang, L.-L. Zhu, and A.~Bonasera,
  ``Number-of-constituent-quark scaling of elliptic flow: a quantitative
  study'', \href{http://dx.doi.org/10.1007/s41365-022-01019-9}{{\em Nucl. Sci.
  Tech.} {\bfseries 33} (Mar., 2022) 37}.
  \url{https://link.springer.com/10.1007/s41365-022-01019-9}.

\bibitem{STAR:2015wza}
{\bfseries STAR} Collaboration, L.~Adamczyk {\em et~al.}, ``{Observation of
  charge asymmetry dependence of pion elliptic flow and the possible chiral
  magnetic wave in heavy-ion collisions}'',
  \href{http://dx.doi.org/10.1103/PhysRevLett.114.252302}{{\em Phys. Rev.
  Lett.} {\bfseries 114} (2015) 252302},
  \href{http://arxiv.org/abs/1504.02175}{{\ttfamily arXiv:1504.02175
  [nucl-ex]}}.

\bibitem{ALICE:2015cjr}
{\bfseries ALICE} Collaboration, J.~Adam {\em et~al.}, ``{Charge-dependent flow
  and the search for the chiral magnetic wave in Pb--Pb collisions at
  $\sqrt{s_{\rm NN}} =$ 2.76 TeV}'',
  \href{http://dx.doi.org/10.1103/PhysRevC.93.044903}{{\em Phys. Rev. C}
  {\bfseries 93} (2016) 044903},
  \href{http://arxiv.org/abs/1512.05739}{{\ttfamily arXiv:1512.05739
  [nucl-ex]}}.

\bibitem{Deng:2012pc}
W.-T. Deng and X.-G. Huang, ``{Event-by-event generation of electromagnetic
  fields in heavy-ion collisions}'',
  \href{http://dx.doi.org/10.1103/PhysRevC.85.044907}{{\em Phys. Rev. C}
  {\bfseries 85} (2012) 044907},
  \href{http://arxiv.org/abs/1201.5108}{{\ttfamily arXiv:1201.5108 [nucl-th]}}.

\bibitem{CMS:2017pah}
{\bfseries CMS} Collaboration, A.~M. Sirunyan {\em et~al.}, ``{Probing the
  chiral magnetic wave in $pPb$ and PbPb collisions at $\sqrt {s_{NN}}$
  =5.02TeV using charge-dependent azimuthal anisotropies}'',
  \href{http://dx.doi.org/10.1103/PhysRevC.100.064908}{{\em Phys. Rev. C}
  {\bfseries 100} (2019) 064908},
  \href{http://arxiv.org/abs/1708.08901}{{\ttfamily arXiv:1708.08901
  [nucl-ex]}}.

\bibitem{Belmont:2016oqp}
R.~Belmont and J.~L. Nagle, ``{Implications of p+Pb measurements on the chiral
  magnetic effect in heavy ion collisions}'',
  \href{http://dx.doi.org/10.1103/PhysRevC.96.024901}{{\em Phys. Rev. C}
  {\bfseries 96} (2017) 024901},
  \href{http://arxiv.org/abs/1610.07964}{{\ttfamily arXiv:1610.07964
  [nucl-th]}}.

\bibitem{STAR:2022hfy}
{\bfseries STAR} Collaboration, M.~I. Abdulhamid {\em et~al.}, ``{Search for
  the chiral magnetic wave using anisotropic flow of identified particles at
  energies available at the BNL Relativistic Heavy Ion Collider}'',
  \href{http://dx.doi.org/10.1103/PhysRevC.108.014908}{{\em Phys. Rev. C}
  {\bfseries 108} (2023) 014908},
  \href{http://arxiv.org/abs/2210.14027}{{\ttfamily arXiv:2210.14027
  [nucl-ex]}}.

\bibitem{Wang:2021nvh}
C.-Z. Wang, W.-Y. Wu, Q.-Y. Shou, G.-L. Ma, Y.-G. Ma, and S.~Zhang,
  ``{Interpreting the charge-dependent flow and constraining the chiral
  magnetic wave with event shape engineering}'',
  \href{http://dx.doi.org/10.1016/j.physletb.2021.136580}{{\em Phys. Lett. B}
  {\bfseries 820} (2021) 136580},
  \href{http://arxiv.org/abs/2104.05551}{{\ttfamily arXiv:2104.05551
  [nucl-th]}}.

\bibitem{Schukraft:2012ah}
J.~Schukraft, A.~Timmins, and S.~A. Voloshin, ``{Ultra-relativistic nuclear
  collisions: event shape engineering}'',
  \href{http://dx.doi.org/10.1016/j.physletb.2013.01.045}{{\em Phys. Lett.}
  {\bfseries B719} (2013) 394--398},
\href{http://arxiv.org/abs/1208.4563}{{\ttfamily arXiv:1208.4563 [nucl-ex]}}.

\bibitem{Bzdak:2013yla}
A.~Bzdak and P.~Bozek, ``{Contributions to the event-by-event charge asymmetry
  dependence for the elliptic flow of $\pi^{+}$ and $\pi^{-}$ in heavy-ion
  collisions}'', \href{http://dx.doi.org/10.1016/j.physletb.2013.08.003}{{\em
  Phys. Lett. B} {\bfseries 726} (2013) 239--243},
  \href{http://arxiv.org/abs/1303.1138}{{\ttfamily arXiv:1303.1138 [nucl-th]}}.

\bibitem{Stephanov:2013tga}
M.~Stephanov and H.-U. Yee, ``{Charged elliptic flow at zero charge
  asymmetry}'', \href{http://dx.doi.org/10.1103/PhysRevC.88.014908}{{\em Phys.
  Rev. C} {\bfseries 88} (2013) 014908},
  \href{http://arxiv.org/abs/1304.6410}{{\ttfamily arXiv:1304.6410 [nucl-th]}}.

\bibitem{Campbell:2013ika}
J.~M. Campbell and M.~A. Lisa, ``{Can baryon stopping explain the breakdown of
  constituent quark scaling and proposed signals of chiral magnetic waves at
  RHIC?}'', \href{http://dx.doi.org/10.1088/1742-6596/446/1/012014}{{\em J.
  Phys. Conf. Ser.} {\bfseries 446} (2013) 012014}.

\bibitem{Voloshin:2014gja}
S.~A. Voloshin and R.~Belmont, ``{Measuring and interpreting charge dependent
  anisotropic flow}'',
  \href{http://dx.doi.org/10.1016/j.nuclphysa.2014.09.030}{{\em Nucl. Phys. A}
  {\bfseries 931} (2014) 992--996},
  \href{http://arxiv.org/abs/1408.0714}{{\ttfamily arXiv:1408.0714 [nucl-ex]}}.

\bibitem{Hatta:2015hca}
Y.~Hatta, A.~Monnai, and B.-W. Xiao, ``{Elliptic flow difference of charged
  pions in heavy-ion collisions}'',
  \href{http://dx.doi.org/10.1016/j.nuclphysa.2015.12.009}{{\em Nucl. Phys. A}
  {\bfseries 947} (2016) 155--160},
  \href{http://arxiv.org/abs/1507.04690}{{\ttfamily arXiv:1507.04690
  [hep-ph]}}.

\bibitem{Hongo:2013cqa}
M.~Hongo, Y.~Hirono, and T.~Hirano, ``{Anomalous-hydrodynamic analysis of
  charge-dependent elliptic flow in heavy-ion collisions}'',
  \href{http://dx.doi.org/10.1016/j.physletb.2017.10.028}{{\em Phys. Lett. B}
  {\bfseries 775} (2017) 266--270},
  \href{http://arxiv.org/abs/1309.2823}{{\ttfamily arXiv:1309.2823 [nucl-th]}}.

\bibitem{Zhao:2019ybo}
X.-L. Zhao, G.-L. Ma, and Y.-G. Ma, ``{Novel mechanism for electric quadrupole
  moment generation in relativistic heavy-ion collisions}'',
  \href{http://dx.doi.org/10.1016/j.physletb.2019.04.002}{{\em Phys. Lett. B}
  {\bfseries 792} (2019) 413--418},
  \href{http://arxiv.org/abs/1901.04156}{{\ttfamily arXiv:1901.04156
  [hep-ph]}}.

\bibitem{Xu:2019pgj}
H.-j. Xu, J.~Zhao, Y.~Feng, and F.~Wang, ``{Complications in the interpretation
  of the charge asymmetry dependent $\pi$ flow for the chiral magnetic wave}'',
  \href{http://dx.doi.org/10.1103/PhysRevC.101.014913}{{\em Phys. Rev. C}
  {\bfseries 101} (2020) 014913},
  \href{http://arxiv.org/abs/1910.02896}{{\ttfamily arXiv:1910.02896
  [nucl-th]}}.

\bibitem{Bass:2000az}
S.~A. Bass, P.~Danielewicz, and S.~Pratt, ``{Clocking hadronization in
  relativistic heavy ion collisions with balance functions}'',
  \href{http://dx.doi.org/10.1103/PhysRevLett.85.2689}{{\em Phys. Rev. Lett.}
  {\bfseries 85} (2000) 2689--2692},
  \href{http://arxiv.org/abs/nucl-th/0005044}{{\ttfamily
  arXiv:nucl-th/0005044}}.

\bibitem{ALICE:2013vrb}
{\bfseries ALICE} Collaboration, B.~Abelev {\em et~al.}, ``{Charge correlations
  using the balance function in Pb--Pb collisions at $\sqrt{s_{\rm NN}}$ = 2.76
  TeV}'', \href{http://dx.doi.org/10.1016/j.physletb.2013.05.039}{{\em Phys.
  Lett. B} {\bfseries 723} (2013) 267--279},
  \href{http://arxiv.org/abs/1301.3756}{{\ttfamily arXiv:1301.3756 [nucl-ex]}}.

\bibitem{Wu:2020wem}
W.-Y. Wu, C.-Z. Wang, Q.-Y. Shou, Y.-G. Ma, and L.~Zheng, ``{Charge-dependent
  transverse momentum and its impact on the search for the chiral magnetic
  wave}'', \href{http://dx.doi.org/10.1103/PhysRevC.103.034906}{{\em Phys. Rev.
  C} {\bfseries 103} (2021) 034906},
  \href{http://arxiv.org/abs/2010.09955}{{\ttfamily arXiv:2010.09955
  [nucl-th]}}.

\bibitem{Ma:2014iva}
G.-L. Ma, ``{Final state effects on charge asymmetry of pion elliptic flow in
  high-energy heavy-ion collisions}'',
  \href{http://dx.doi.org/10.1016/j.physletb.2014.06.074}{{\em Phys. Lett. B}
  {\bfseries 735} (2014) 383--386},
  \href{http://arxiv.org/abs/1401.6502}{{\ttfamily arXiv:1401.6502 [nucl-th]}}.

\bibitem{Zhou:2018rkh}
W.-H. Zhou and J.~Xu, ``{Simulating the Chiral Magnetic Wave in a Box
  System}'', \href{http://dx.doi.org/10.1103/PhysRevC.98.044904}{{\em Phys.
  Rev. C} {\bfseries 98} (2018) 044904},
  \href{http://arxiv.org/abs/1810.01030}{{\ttfamily arXiv:1810.01030
  [nucl-th]}}.

\bibitem{Han:2019fce}
Z.-Z. Han and J.~Xu, ``{Charge asymmetry dependence of the elliptic flow
  splitting in relativistic heavy-ion collisions}'',
  \href{http://dx.doi.org/10.1103/PhysRevC.99.044915}{{\em Phys. Rev. C}
  {\bfseries 99} (2019) 044915},
  \href{http://arxiv.org/abs/1904.03544}{{\ttfamily arXiv:1904.03544
  [nucl-th]}}.

\bibitem{Shen:2019puh}
D.~Shen, J.~Chen, G.~Ma, Y.-G. Ma, Q.~Shou, S.~Zhang, and C.~Zhong, ``{Charge
  asymmetry dependence of flow and a novel correlator to detect the chiral
  magnetic wave in a multiphase transport model}'',
  \href{http://dx.doi.org/10.1103/PhysRevC.100.064907}{{\em Phys. Rev. C}
  {\bfseries 100} (2019) 064907},
  \href{http://arxiv.org/abs/1911.00839}{{\ttfamily arXiv:1911.00839
  [hep-ph]}}.

\bibitem{Magdy:2020xqs}
N.~Magdy, M.-W. Nie, L.~Huang, G.-L. Ma, and R.~A. Lacey, ``{An extended
  $R^{(2)}_{\Psi_m}(\Delta S_2)$ correlator for detecting and characterizing
  the Chiral Magnetic Wave}'',
  \href{http://dx.doi.org/10.1016/j.physletb.2020.135986}{{\em Phys. Lett. B}
  {\bfseries 811} (2020) 135986},
  \href{http://arxiv.org/abs/2003.02396}{{\ttfamily arXiv:2003.02396
  [nucl-ex]}}.

\bibitem{Shi:2017cpu}
S.~Shi, Y.~Jiang, E.~Lilleskov, and J.~Liao, ``{Anomalous Chiral Transport in
  Heavy Ion Collisions from Anomalous-Viscous Fluid Dynamics}'',
  \href{http://dx.doi.org/10.1016/j.aop.2018.04.026}{{\em Annals Phys.}
  {\bfseries 394} (2018) 50--72},
  \href{http://arxiv.org/abs/1711.02496}{{\ttfamily arXiv:1711.02496
  [nucl-th]}}.

\bibitem{wu_global_2023}
W.-Y. Wu {\em et~al.}, ``{Global constraint on the magnitude of anomalous
  chiral effects in heavy-ion collisions}'',
  \href{http://dx.doi.org/10.1103/PhysRevC.107.L031902}{{\em Phys. Rev. C}
  {\bfseries 107} (2023) L031902},
  \href{http://arxiv.org/abs/2211.15446}{{\ttfamily arXiv:2211.15446
  [nucl-th]}}.

\bibitem{Aamodt:2008zz}
{\bfseries ALICE} Collaboration, K.~Aamodt {\em et~al.}, ``{The ALICE
  experiment at the CERN LHC}'',
\href{http://dx.doi.org/10.1088/1748-0221/3/08/S08002}{{\em JINST} {\bfseries
  3} (2008) S08002}.

\bibitem{Abelev:2014ffa}
{\bfseries ALICE} Collaboration, B.~B. Abelev {\em et~al.}, ``{Performance of
  the ALICE Experiment at the CERN LHC}'',
  \href{http://dx.doi.org/10.1142/S0217751X14300440}{{\em Int. J. Mod. Phys.}
  {\bfseries A29} (2014) 1430044},
\href{http://arxiv.org/abs/1402.4476}{{\ttfamily arXiv:1402.4476 [nucl-ex]}}.

\bibitem{Alme:2010ke}
J.~Alme {\em et~al.}, ``{The ALICE TPC, a large 3-dimensional tracking device
  with fast readout for ultra-high multiplicity events}'',
  \href{http://dx.doi.org/10.1016/j.nima.2010.04.042}{{\em Nucl. Instrum.
  Meth.} {\bfseries A622} (2010) 316--367},
\href{http://arxiv.org/abs/1001.1950}{{\ttfamily arXiv:1001.1950
  [physics.ins-det]}}.

\bibitem{ALICE:2000xcm}
{\bfseries ALICE} Collaboration, G.~Dellacasa {\em et~al.}, ``{ALICE technical
  design report of the time-of-flight system (TOF)}'', {\em
  \href{http://cds.cern.ch/record/430132}{CERN-LHCC-2000-012}} .

\bibitem{Abbas:2013taa}
{\bfseries ALICE} Collaboration, E.~Abbas {\em et~al.}, ``{Performance of the
  ALICE VZERO system}'',
  \href{http://dx.doi.org/10.1088/1748-0221/8/10/P10016}{{\em JINST} {\bfseries
  8} (2013) P10016},
\href{http://arxiv.org/abs/1306.3130}{{\ttfamily arXiv:1306.3130 [nucl-ex]}}.

\bibitem{ALICE:2020siw}
{\bfseries ALICE} Collaboration, S.~Acharya {\em et~al.}, ``{Constraining the
  Chiral Magnetic Effect with charge-dependent azimuthal correlations in Pb--Pb
  collisions at $ \sqrt{s_{\mathrm{NN}}} $ = 2.76 and 5.02 TeV}'',
  \href{http://dx.doi.org/10.1007/JHEP09(2020)160}{{\em JHEP} {\bfseries 09}
  (2020) 160}, \href{http://arxiv.org/abs/2005.14640}{{\ttfamily
  arXiv:2005.14640 [nucl-ex]}}.

\bibitem{ALICE:2013hur}
{\bfseries ALICE} Collaboration, B.~Abelev {\em et~al.}, ``{Centrality
  determination of Pb-Pb collisions at $\sqrt{s_{\rm{NN}}}$ = 2.76 TeV with
  ALICE}'', \href{http://dx.doi.org/10.1103/PhysRevC.88.044909}{{\em Phys. Rev.
  C} {\bfseries 88} (2013) 044909},
  \href{http://arxiv.org/abs/1301.4361}{{\ttfamily arXiv:1301.4361 [nucl-ex]}}.

\bibitem{Gyulassy:1994ew}
M.~Gyulassy and X.-N. Wang, ``{HIJING 1.0: A Monte Carlo program for parton and
  particle production in high-energy hadronic and nuclear collisions}'',
  \href{http://dx.doi.org/10.1016/0010-4655(94)90057-4}{{\em Comput. Phys.
  Commun.} {\bfseries 83} (1994) 307},
  \href{http://arxiv.org/abs/nucl-th/9502021}{{\ttfamily
  arXiv:nucl-th/9502021}}.

\bibitem{Brun:1987ma}
R.~Brun, F.~Bruyant, F.~Carminati, S.~Giani, M.~Maire, A.~McPherson,
  G.~Patrick, and L.~Urban, ``{GEANT Detector Description and Simulation
  Tool}'', \href{http://dx.doi.org/10.17181/CERN.MUHF.DMJ1}{{\em CERN Program
  Library, CERN, Geneva} (1993) }.

\bibitem{Bilandzic:2010jr}
A.~Bilandzic, R.~Snellings, and S.~Voloshin, ``{Flow analysis with cumulants:
  Direct calculations}'',
  \href{http://dx.doi.org/10.1103/PhysRevC.83.044913}{{\em Phys. Rev. C}
  {\bfseries 83} (2011) 044913},
  \href{http://arxiv.org/abs/1010.0233}{{\ttfamily arXiv:1010.0233 [nucl-ex]}}.

\bibitem{Zhou:2015iba}
Y.~Zhou, X.~Zhu, P.~Li, and H.~Song, ``{Investigation of possible hadronic flow
  in $\sqrt{s_{NN}} = 5.02$ TeV $p-Pb$ collisions}'',
  \href{http://dx.doi.org/10.1103/PhysRevC.91.064908}{{\em Phys. Rev. C}
  {\bfseries 91} (2015) 064908},
  \href{http://arxiv.org/abs/1503.06986}{{\ttfamily arXiv:1503.06986
  [nucl-th]}}.

\bibitem{Barlow:2002yb}
R.~Barlow, ``{Systematic errors: Facts and fictions}'', in {\em {Advanced
  Statistical Techniques in Particle Physics. Proceedings, Conference, Durham,
  UK, March 18-22, 2002}}, pp.~134--144.
\newblock 2002.
\newblock
\href{http://arxiv.org/abs/hep-ex/0207026}{{\ttfamily arXiv:hep-ex/0207026
  [hep-ex]}}.
\newblock

\bibitem{Xu:2020sln}
H.-j. Xu, J.~Zhao, Y.~Feng, and F.~Wang, ``{Importance of non-flow background
  on the chiral magnetic wave search}'',
  \href{http://dx.doi.org/10.1016/j.nuclphysa.2020.121770}{{\em Nucl. Phys. A}
  {\bfseries 1005} (2021) 121770},
  \href{http://arxiv.org/abs/2002.05220}{{\ttfamily arXiv:2002.05220
  [nucl-th]}}.

\bibitem{ALICE:2016kpq}
{\bfseries ALICE} Collaboration, J.~Adam {\em et~al.}, ``{Correlated
  event-by-event fluctuations of flow harmonics in Pb--Pb collisions at
  $\sqrt{s_{_{\rm NN}}}=2.76$ TeV}'',
  \href{http://dx.doi.org/10.1103/PhysRevLett.117.182301}{{\em Phys. Rev.
  Lett.} {\bfseries 117} (2016) 182301},
  \href{http://arxiv.org/abs/1604.07663}{{\ttfamily arXiv:1604.07663
  [nucl-ex]}}.

\bibitem{ALICE:2011ab}
{\bfseries ALICE} Collaboration, K.~Aamodt {\em et~al.}, ``{Higher harmonic
  anisotropic flow measurements of charged particles in Pb--Pb collisions at
  $\sqrt{s_{\rm NN}}$=2.76 TeV}'',
  \href{http://dx.doi.org/10.1103/PhysRevLett.107.032301}{{\em Phys. Rev.
  Lett.} {\bfseries 107} (2011) 032301},
  \href{http://arxiv.org/abs/1105.3865}{{\ttfamily arXiv:1105.3865 [nucl-ex]}}.

\bibitem{ATLAS:2014ndd}
{\bfseries ATLAS} Collaboration, G.~Aad {\em et~al.}, ``{Measurement of
  event-plane correlations in $\sqrt{s_{NN}}=2.76$ TeV lead-lead collisions
  with the ATLAS detector}'',
  \href{http://dx.doi.org/10.1103/PhysRevC.90.024905}{{\em Phys. Rev. C}
  {\bfseries 90} (2014) 024905},
  \href{http://arxiv.org/abs/1403.0489}{{\ttfamily arXiv:1403.0489 [hep-ex]}}.

\bibitem{Hatta:2016czn}
Y.~Hatta, ``{Analytic approaches to relativistic hydrodynamics}'',
  \href{http://dx.doi.org/10.1016/j.nuclphysa.2016.02.004}{{\em Nucl. Phys. A}
  {\bfseries 956} (2016) 152--159},
  \href{http://arxiv.org/abs/1601.04128}{{\ttfamily arXiv:1601.04128
  [hep-ph]}}.

\end{thebibliography}\endgroup

\newpage
\appendix

%
%

\section{The ALICE Collaboration}
\label{app:collab}
\begin{flushleft} 
\small

S.~Acharya\,\orcidlink{0000-0002-9213-5329}\,$^{\rm 128}$, 
D.~Adamov\'{a}\,\orcidlink{0000-0002-0504-7428}\,$^{\rm 87}$, 
G.~Aglieri Rinella\,\orcidlink{0000-0002-9611-3696}\,$^{\rm 33}$, 
M.~Agnello\,\orcidlink{0000-0002-0760-5075}\,$^{\rm 30}$, 
N.~Agrawal\,\orcidlink{0000-0003-0348-9836}\,$^{\rm 52}$, 
Z.~Ahammed\,\orcidlink{0000-0001-5241-7412}\,$^{\rm 136}$, 
S.~Ahmad\,\orcidlink{0000-0003-0497-5705}\,$^{\rm 16}$, 
S.U.~Ahn\,\orcidlink{0000-0001-8847-489X}\,$^{\rm 72}$, 
I.~Ahuja\,\orcidlink{0000-0002-4417-1392}\,$^{\rm 38}$, 
A.~Akindinov\,\orcidlink{0000-0002-7388-3022}\,$^{\rm 142}$
% ...

\section*{Affiliation Notes}

$^{\rm I}$ Deceased\\
$^{\rm II}$ Also at: Max-Planck-Institut fur Physik, Munich, Germany\\
$^{\rm III}$ Also at: Italian National Agency for New Technologies, Energy and Sustainable Economic Development (ENEA), Bologna, Italy\\
$^{\rm IV}$ Also at: Dipartimento DET del Politecnico di Torino, Turin, Italy\\
$^{\rm V}$ Also at: Department of Applied Physics, Aligarh Muslim University, Aligarh, India\\
$^{\rm VI}$ Also at: Institute of Theoretical Physics, University of Wroclaw, Poland\\
$^{\rm VII}$ Also at: An institution covered by a cooperation agreement with CERN\\

\section*{Collaboration Institutes}

$^{1}$ A.I. Alikhanyan National Science Laboratory (Yerevan Physics Institute) Foundation, Yerevan, Armenia\\
$^{2}$ AGH University of Krakow, Cracow, Poland\\
$^{3}$ Bogolyubov Institute for Theoretical Physics, National Academy of Sciences of Ukraine, Kiev, Ukraine\\
$^{4}$ Bose Institute, Department of Physics  and Centre for Astroparticle Physics and Space Science (CAPSS), Kolkata, India\\
$^{5}$ California Polytechnic State University, San Luis Obispo, California, United States\\
$^{6}$ Central China Normal University, Wuhan, China\\
$^{7}$ Centro de Aplicaciones Tecnol\'{o}gicas y Desarrollo Nuclear (CEADEN), Havana, Cuba\\
$^{8}$ Centro de Investigaci\'{o}n y de Estudios Avanzados (CINVESTAV), Mexico City and M\'{e}rida, Mexico\\
$^{9}$ Chicago State University, Chicago, Illinois, United States\\
$^{10}$ China Institute of Atomic Energy, Beijing, China\\
$^{11}$ China University of Geosciences, Wuhan, China\\
$^{12}$ Chungbuk National University, Cheongju, Republic of Korea\\
$^{13}$ Comenius University Bratislava, Faculty of Mathematics, Physics and Informatics, Bratislava, Slovak Republic\\
$^{14}$ COMSATS University Islamabad, Islamabad, Pakistan\\
$^{15}$ Creighton University, Omaha, Nebraska, United States\\
$^{16}$ Department of Physics, Aligarh Muslim University, Aligarh, India\\
$^{17}$ Department of Physics, Pusan National University, Pusan, Republic of Korea\\
$^{18}$ Department of Physics, Sejong University, Seoul, Republic of Korea\\
$^{19}$ Department of Physics, University of California, Berkeley, California, United States\\
$^{20}$ Department of Physics, University of Oslo, Oslo, Norway\\
$^{21}$ Department of Physics and Technology, University of Bergen, Bergen, Norway\\
$^{22}$ Dipartimento di Fisica, Universit\`{a} di Pavia, Pavia, Italy\\
$^{23}$ Dipartimento di Fisica dell'Universit\`{a} and Sezione INFN, Cagliari, Italy\\
$^{24}$ Dipartimento di Fisica dell'Universit\`{a} and Sezione INFN, Trieste, Italy\\
$^{25}$ Dipartimento di Fisica dell'Universit\`{a} and Sezione INFN, Turin, Italy\\
$^{26}$ Dipartimento di Fisica e Astronomia dell'Universit\`{a} and Sezione INFN, Bologna, Italy\\
$^{27}$ Dipartimento di Fisica e Astronomia dell'Universit\`{a} and Sezione INFN, Catania, Italy\\
$^{28}$ Dipartimento di Fisica e Astronomia dell'Universit\`{a} and Sezione INFN, Padova, Italy\\
$^{29}$ Dipartimento di Fisica `E.R.~Caianiello' dell'Universit\`{a} and Gruppo Collegato INFN, Salerno, Italy\\
$^{30}$ Dipartimento DISAT del Politecnico and Sezione INFN, Turin, Italy\\
$^{31}$ Dipartimento di Scienze MIFT, Universit\`{a} di Messina, Messina, Italy\\
$^{32}$ Dipartimento Interateneo di Fisica `M.~Merlin' and Sezione INFN, Bari, Italy\\
$^{33}$ European Organization for Nuclear Research (CERN), Geneva, Switzerland\\
$^{34}$ Faculty of Electrical Engineering, Mechanical Engineering and Naval Architecture, University of Split, Split, Croatia\\
$^{35}$ Faculty of Engineering and Science, Western Norway University of Applied Sciences, Bergen, Norway\\
$^{36}$ Faculty of Nuclear Sciences and Physical Engineering, Czech Technical University in Prague, Prague, Czech Republic\\
$^{37}$ Faculty of Physics, Sofia University, Sofia, Bulgaria\\
$^{38}$ Faculty of Science, P.J.~\v{S}af\'{a}rik University, Ko\v{s}ice, Slovak Republic\\
$^{39}$ Frankfurt Institute for Advanced Studies, Johann Wolfgang Goethe-Universit\"{a}t Frankfurt, Frankfurt, Germany\\
$^{40}$ Fudan University, Shanghai, China\\
$^{41}$ Gangneung-Wonju National University, Gangneung, Republic of Korea\\
$^{42}$ Gauhati University, Department of Physics, Guwahati, India\\
$^{43}$ Helmholtz-Institut f\"{u}r Strahlen- und Kernphysik, Rheinische Friedrich-Wilhelms-Universit\"{a}t Bonn, Bonn, Germany\\
$^{44}$ Helsinki Institute of Physics (HIP), Helsinki, Finland\\
$^{45}$ High Energy Physics Group,  Universidad Aut\'{o}noma de Puebla, Puebla, Mexico\\
$^{46}$ Horia Hulubei National Institute of Physics and Nuclear Engineering, Bucharest, Romania\\
$^{47}$ HUN-REN Wigner Research Centre for Physics, Budapest, Hungary\\
$^{48}$ Indian Institute of Technology Bombay (IIT), Mumbai, India\\
$^{49}$ Indian Institute of Technology Indore, Indore, India\\
$^{50}$ INFN, Laboratori Nazionali di Frascati, Frascati, Italy\\
$^{51}$ INFN, Sezione di Bari, Bari, Italy\\
$^{52}$ INFN, Sezione di Bologna, Bologna, Italy\\
$^{53}$ INFN, Sezione di Cagliari, Cagliari, Italy\\
$^{54}$ INFN, Sezione di Catania, Catania, Italy\\
$^{55}$ INFN, Sezione di Padova, Padova, Italy\\
$^{56}$ INFN, Sezione di Pavia, Pavia, Italy\\
$^{57}$ INFN, Sezione di Torino, Turin, Italy\\
$^{58}$ INFN, Sezione di Trieste, Trieste, Italy\\
$^{59}$ Inha University, Incheon, Republic of Korea\\
$^{60}$ Institute for Gravitational and Subatomic Physics (GRASP), Utrecht University/Nikhef, Utrecht, Netherlands\\
$^{61}$ Institute of Experimental Physics, Slovak Academy of Sciences, Ko\v{s}ice, Slovak Republic\\
$^{62}$ Institute of Physics, Homi Bhabha National Institute, Bhubaneswar, India\\
$^{63}$ Institute of Physics of the Czech Academy of Sciences, Prague, Czech Republic\\
$^{64}$ Institute of Space Science (ISS), Bucharest, Romania\\
$^{65}$ Institut f\"{u}r Kernphysik, Johann Wolfgang Goethe-Universit\"{a}t Frankfurt, Frankfurt, Germany\\
$^{66}$ Instituto de Ciencias Nucleares, Universidad Nacional Aut\'{o}noma de M\'{e}xico, Mexico City, Mexico\\
$^{67}$ Instituto de F\'{i}sica, Universidade Federal do Rio Grande do Sul (UFRGS), Porto Alegre, Brazil\\
$^{68}$ Instituto de F\'{\i}sica, Universidad Nacional Aut\'{o}noma de M\'{e}xico, Mexico City, Mexico\\
$^{69}$ iThemba LABS, National Research Foundation, Somerset West, South Africa\\
$^{70}$ Jeonbuk National University, Jeonju, Republic of Korea\\
$^{71}$ Johann-Wolfgang-Goethe Universit\"{a}t Frankfurt Institut f\"{u}r Informatik, Fachbereich Informatik und Mathematik, Frankfurt, Germany\\
$^{72}$ Korea Institute of Science and Technology Information, Daejeon, Republic of Korea\\
$^{73}$ KTO Karatay University, Konya, Turkey\\
$^{74}$ Laboratoire de Physique Subatomique et de Cosmologie, Universit\'{e} Grenoble-Alpes, CNRS-IN2P3, Grenoble, France\\
$^{75}$ Lawrence Berkeley National Laboratory, Berkeley, California, United States\\
$^{76}$ Lund University Department of Physics, Division of Particle Physics, Lund, Sweden\\
$^{77}$ Nagasaki Institute of Applied Science, Nagasaki, Japan\\
$^{78}$ Nara Women{'}s University (NWU), Nara, Japan\\
$^{79}$ National and Kapodistrian University of Athens, School of Science, Department of Physics , Athens, Greece\\
$^{80}$ National Centre for Nuclear Research, Warsaw, Poland\\
$^{81}$ National Institute of Science Education and Research, Homi Bhabha National Institute, Jatni, India\\
$^{82}$ National Nuclear Research Center, Baku, Azerbaijan\\
$^{83}$ National Research and Innovation Agency - BRIN, Jakarta, Indonesia\\
$^{84}$ Niels Bohr Institute, University of Copenhagen, Copenhagen, Denmark\\
$^{85}$ Nikhef, National institute for subatomic physics, Amsterdam, Netherlands\\
$^{86}$ Nuclear Physics Group, STFC Daresbury Laboratory, Daresbury, United Kingdom\\
$^{87}$ Nuclear Physics Institute of the Czech Academy of Sciences, Husinec-\v{R}e\v{z}, Czech Republic\\
$^{88}$ Oak Ridge National Laboratory, Oak Ridge, Tennessee, United States\\
$^{89}$ Ohio State University, Columbus, Ohio, United States\\
$^{90}$ Physics department, Faculty of science, University of Zagreb, Zagreb, Croatia\\
$^{91}$ Physics Department, Panjab University, Chandigarh, India\\
$^{92}$ Physics Department, University of Jammu, Jammu, India\\
$^{93}$ Physics Program and International Institute for Sustainability with Knotted Chiral Meta Matter (SKCM2), Hiroshima University, Hiroshima, Japan\\
$^{94}$ Physikalisches Institut, Eberhard-Karls-Universit\"{a}t T\"{u}bingen, T\"{u}bingen, Germany\\
$^{95}$ Physikalisches Institut, Ruprecht-Karls-Universit\"{a}t Heidelberg, Heidelberg, Germany\\
$^{96}$ Physik Department, Technische Universit\"{a}t M\"{u}nchen, Munich, Germany\\
$^{97}$ Politecnico di Bari and Sezione INFN, Bari, Italy\\
$^{98}$ Research Division and ExtreMe Matter Institute EMMI, GSI Helmholtzzentrum f\"ur Schwerionenforschung GmbH, Darmstadt, Germany\\
$^{99}$ Saga University, Saga, Japan\\
$^{100}$ Saha Institute of Nuclear Physics, Homi Bhabha National Institute, Kolkata, India\\
$^{101}$ School of Physics and Astronomy, University of Birmingham, Birmingham, United Kingdom\\
$^{102}$ Secci\'{o}n F\'{\i}sica, Departamento de Ciencias, Pontificia Universidad Cat\'{o}lica del Per\'{u}, Lima, Peru\\
$^{103}$ Stefan Meyer Institut f\"{u}r Subatomare Physik (SMI), Vienna, Austria\\
$^{104}$ SUBATECH, IMT Atlantique, Nantes Universit\'{e}, CNRS-IN2P3, Nantes, France\\
$^{105}$ Sungkyunkwan University, Suwon City, Republic of Korea\\
$^{106}$ Suranaree University of Technology, Nakhon Ratchasima, Thailand\\
$^{107}$ Technical University of Ko\v{s}ice, Ko\v{s}ice, Slovak Republic\\
$^{108}$ The Henryk Niewodniczanski Institute of Nuclear Physics, Polish Academy of Sciences, Cracow, Poland\\
$^{109}$ The University of Texas at Austin, Austin, Texas, United States\\
$^{110}$ Universidad Aut\'{o}noma de Sinaloa, Culiac\'{a}n, Mexico\\
$^{111}$ Universidade de S\~{a}o Paulo (USP), S\~{a}o Paulo, Brazil\\
$^{112}$ Universidade Estadual de Campinas (UNICAMP), Campinas, Brazil\\
$^{113}$ Universidade Federal do ABC, Santo Andre, Brazil\\
$^{114}$ Universitatea Nationala de Stiinta si Tehnologie Politehnica Bucuresti, Bucharest, Romania\\
$^{115}$ University of Cape Town, Cape Town, South Africa\\
$^{116}$ University of Derby, Derby, United Kingdom\\
$^{117}$ University of Houston, Houston, Texas, United States\\
$^{118}$ University of Jyv\"{a}skyl\"{a}, Jyv\"{a}skyl\"{a}, Finland\\
$^{119}$ University of Kansas, Lawrence, Kansas, United States\\
$^{120}$ University of Liverpool, Liverpool, United Kingdom\\
$^{121}$ University of Science and Technology of China, Hefei, China\\
$^{122}$ University of South-Eastern Norway, Kongsberg, Norway\\
$^{123}$ University of Tennessee, Knoxville, Tennessee, United States\\
$^{124}$ University of the Witwatersrand, Johannesburg, South Africa\\
$^{125}$ University of Tokyo, Tokyo, Japan\\
$^{126}$ University of Tsukuba, Tsukuba, Japan\\
$^{127}$ Universit\"{a}t M\"{u}nster, Institut f\"{u}r Kernphysik, M\"{u}nster, Germany\\
$^{128}$ Universit\'{e} Clermont Auvergne, CNRS/IN2P3, LPC, Clermont-Ferrand, France\\
$^{129}$ Universit\'{e} de Lyon, CNRS/IN2P3, Institut de Physique des 2 Infinis de Lyon, Lyon, France\\
$^{130}$ Universit\'{e} de Strasbourg, CNRS, IPHC UMR 7178, F-67000 Strasbourg, France, Strasbourg, France\\
$^{131}$ Universit\'{e} Paris-Saclay, Centre d'Etudes de Saclay (CEA), IRFU, D\'{e}partment de Physique Nucl\'{e}aire (DPhN), Saclay, France\\
$^{132}$ Universit\'{e}  Paris-Saclay, CNRS/IN2P3, IJCLab, Orsay, France\\
$^{133}$ Universit\`{a} degli Studi di Foggia, Foggia, Italy\\
$^{134}$ Universit\`{a} del Piemonte Orientale, Vercelli, Italy\\
$^{135}$ Universit\`{a} di Brescia, Brescia, Italy\\
$^{136}$ Variable Energy Cyclotron Centre, Homi Bhabha National Institute, Kolkata, India\\
$^{137}$ Warsaw University of Technology, Warsaw, Poland\\
$^{138}$ Wayne State University, Detroit, Michigan, United States\\
$^{139}$ Yale University, New Haven, Connecticut, United States\\
$^{140}$ Yonsei University, Seoul, Republic of Korea\\
$^{141}$  Zentrum  f\"{u}r Technologie und Transfer (ZTT), Worms, Germany\\
$^{142}$ Affiliated with an institute covered by a cooperation agreement with CERN\\
$^{143}$ Affiliated with an international laboratory covered by a cooperation agreement with CERN.\\

\end{flushleft} 

\end{document}